\RequirePackage{fixltx2e}
\documentclass[11pt,a4paper]{article}
\pdfoutput=1
\usepackage[utf8]{inputenc}
\usepackage{jheppub}
\hypersetup{unicode=true,bookmarksopen=true}

\usepackage{subcaption}

\usepackage{amsthm}
\usepackage{mathtools}
\usepackage{lmodern}
\usepackage[T1]{fontenc}
\usepackage{bbm}
\DeclareMathAlphabet{\mathbfi}{OML}{cmm}{b}{it}

\usepackage[nointegrals]{wasysym}

\let\originalleft\left
\let\originalright\right
\renewcommand{\left}{\mathopen{}\mathclose\bgroup\originalleft}
\renewcommand{\right}{\aftergroup\egroup\originalright}
\makeatletter
\newcommand{\biggg}{\bBigg@\thr@@}
\newcommand{\Biggg}{\bBigg@{3.5}}
\makeatother

\makeatletter
\newenvironment{equations}[1][]{\subequations\ifx\relax#1\relax\else\label{#1}\fi\align\ignorespaces}{\endalign\ignorespacesafterend\endsubequations}
\def\@spliteq#1{\begin{equation}\begin{split}#1\end{split}\end{equation}}
\def\@spliteqstar#1{\begin{equation*}\begin{split}#1\end{split}\end{equation*}}
\def\splitequation{\collect@body\@spliteq}
\expandafter\def\csname splitequation*\endcsname{\collect@body\@spliteqstar}

\expandafter\def\csname endsplitequation*\endcsname{\ignorespacesafterend}
\makeatother

\renewcommand{\vec}[1]{{\ifnum9<1#1\mathbf{#1}\else\ifcat\noexpand#1\relax\boldsymbol{#1}\else\mathbfi{#1}\fi\fi}}
\newcommand{\mathe}{\mathrm{e}}
\newcommand{\mathi}{\mathrm{i}}
\let\oldre\Re
\let\oldim\Im
\renewcommand{\Re}{\oldre\mathfrak{e}\,}
\renewcommand{\Im}{\oldim\mathfrak{m}\,}
\newcommand{\total}{\mathop{}\!\mathrm{d}}
\newcommand{\abs}[1]{{\left\lvert{#1}\right\rvert}}
\newcommand{\norm}[1]{{\left\lVert{#1}\right\rVert}}

\newcommand{\1}{\mathbbm{1}}
\newcommand{\tr}{\operatorname{tr}}

\newcommand{\eqend}[1]{\,#1}
\newcommand{\bigo}[1]{\mathcal{O}\left({#1}\right)}

\newcommand{\bra}[1]{\left\langle{#1}\right\vert}
\newcommand{\ket}[1]{\left\vert{#1}\right\rangle}

\newcommand{\normord}[1]{\mathopen{:}{#1}\mathclose{:}}
\newcommand{\supp}{\operatorname{supp}}
\newcommand{\pf}{\mathcal{P}\!f}

\makeatletter
\newcommand{\dhat}[1]{%
\begingroup%
  \let\macc@kerna\z@%
  \let\macc@kernb\z@%
  \let\macc@nucleus\@empty%
  \hat{\mathchoice%
    {\raisebox{.4ex}{\vphantom{\ensuremath{\displaystyle #1}}}}%
    {\raisebox{.4ex}{\vphantom{\ensuremath{\textstyle #1}}}}%
    {\raisebox{.32ex}{\vphantom{\ensuremath{\scriptstyle #1}}}}%
    {\raisebox{.28ex}{\vphantom{\ensuremath{\scriptscriptstyle #1}}}}%
    \smash{\hat{#1}}}%
\endgroup%
}
\makeatother

\makeatletter
\DeclareRobustCommand*{\citeDLMFeq}{\hyper@normalise\citeDLMFeq@}
\def\citeDLMFeq@#1#2{\cite[\hyper@linkurl{Eq.~#1.#2}{https://dlmf.nist.gov/#1.E#2}]{DLMF}}
\makeatother

\allowdisplaybreaks

\makeatletter
\gdef\@fpheader{\strut}
\makeatother

\makeatletter
\g@addto@macro\@floatboxreset\centering
\makeatother

\newtheorem{theorem}{Theorem}
\newtheorem{lemma}{Lemma}
\newtheorem{corollary}{Corollary}

\begin{document}

\title{Modular Hamiltonian and modular flow of massless fermions on a cylinder}

\author[a]{Daniela Cadamuro,}
\author[a]{Markus B. Fröb}
\author[b,c]{and Guillem Pérez-Nadal}

\affiliation[a]{Institut f{\"u}r Theoretische Physik, Universit{\"a}t Leipzig, Br{\"u}derstra{\ss}e 16, 04103 Leipzig, Germany}
\affiliation[b]{Departamento de F{\'\i}sica, Facultad de Ciencias Exactas y Naturales, Universidad de Buenos Aires, 1428 Buenos Aires, Argentina}
\affiliation[c]{CONICET -- Instituto de F{\'\i}sica de Buenos Aires (IFIBA), Universidad de Buenos Aires, 1428 Buenos Aires, Argentina}

\emailAdd{cadamuro@itp.uni-leipzig.de}
\emailAdd{mfroeb@itp.uni-leipzig.de}
\emailAdd{guillem@df.uba.ar}

\abstract{We determine explicitly the modular flow and the modular Hamiltonian for massless free fermions in diamonds on a cylinder in 1+1 dimensions. We consider both periodic and antiperiodic boundary conditions, the ground state in the antiperiodic case and the most general family of quasi-free zero-energy ground states in the periodic case, which depend on four parameters and are generally mixed. While for the antiperiodic ground state and one periodic ground state (the maximally mixed zero-temperature state) the modular data is known, our results for the generic ground state in the periodic case are completely new. We find that generically both the modular flow and the modular Hamiltonian are non-local, and we show that in the parametric limit where the state becomes pure the modular data becomes local. Moreover, even in the local case the modular flow generically mixes the two chiralities. This kind of behavior has not been observed previously.}


\maketitle

\section{Introduction}
\label{sec:intro}

The Tomita--Takesaki theory of modular flows of von Neumann algebras~\cite{takesaki1970} has played an increasingly important role recently in understanding the entanglement structure of quantum field theories, see for example Refs.~\cite{witten2018,bostelmanncadamurodelvecchio2022,ciollilongoranalloruzzi2022,garbarzpalau2023,digiulioerdmenger2023} and references therein. The theory originates in the study of operator algebras~\cite{bratellirobinson1,bratellirobinson2,takesaki2002a,takesaki2002b}, and in fact can be used to classify von Neumann algebras. In relativistic quantum field theory, the local von Neumann algebras $\mathfrak{A}(\mathcal{R})$ that are associated to spacetime regions $\mathcal{R}$ are generically isomorphic to the unique hyperfinite type III${}_1$ factor~\cite{fredenhagen1985,buchholzdantonifredenhagen1987,buchholzverch1995}, see also Ref.~\cite{yngvason2005}. In this case, the modular flow for any faithful normal state is an outer automorphism of $\mathfrak{A}$, and the spectrum of the modular Hamiltonian, the generator of the flow, is the full real line.

For type III algebras, density matrices don't exist as operators and all non-zero traces are infinite. Therefore, standard measures of entanglement that are defined for type I algebras, such as the von Neumann entropy $S(\rho) = - \tr\left( \rho \ln \rho \right)$ evaluated for a reduced density matrix $\rho$, are divergent in quantum field theory. On the other hand, quantities such as the relative entropy (or Kullback--Leibler divergence) $S(\rho\Vert\sigma) = \tr\left( \rho \ln \rho - \rho \ln \sigma \right)$ that measures the distinguishability of two states remain finite also for quantum field theories. This is seen by relating the relative entropy to the relative modular Hamiltonian via the Araki--Uhlmann formula~\cite{araki1976,uhlmann1977}
\begin{equation}
\label{eq:relative_entropy_def}
S\left( \Phi \Vert \Psi \right) = - \left( \Phi, \ln \Delta_{\Phi\vert\Psi} \Phi \right) \eqend{,}
\end{equation}
where $\Phi$ and $\Psi$ are the vector representatives of two faithful normal states, and $\ln \Delta_{\Phi\vert\Psi}$ is the relative modular Hamiltonian, the generator of the relative modular flow defined for the two states. For a type I algebra, the definition~\eqref{eq:relative_entropy_def} is equal to the one in terms of density matrices (see for example~\cite[Sec.~IV]{witten2018}), but it clearly also makes sense for a type III algebra.

To compute relative entropies via the Araki--Uhlmann formula~\eqref{eq:relative_entropy_def}, it is thus necessary to know the relative modular Hamiltonian, or alternatively the full (relative) modular flow. They depend on both the von Neumann algebra under study (which in quantum field theory is the algebra $\mathfrak{A}(\mathcal{R})$ of bounded operators associated to a spacetime region $\mathcal{R}$) and the pair of states. For states which are obtained as unitary excitations of another state (for example the vacuum), the relative modular flow has a simple relation to the flow of this other state. It is thus enough to concentrate on the latter one, which is the approach that is considered in practice. In the simplest cases, the flow is local and geometric. For general quantum field theories, this happens for $\mathcal{R}$ the Rindler wedge and for the Minkowski vacuum state (the Bisognano--Wichmann theorem~\cite{bisognanowichmann1976}), for $\mathcal{R}$ equal to the static patch in de Sitter spacetime (which is the intersection of a wedge in the embedding Minkowski spacetime with the de Sitter hyperboloid) and for the de Sitter (Bunch--Davies or Euclidean) vacuum~\cite{figarihoeghkrohnnappi1975,gibbonshawking1977,borchersbuchholz1999}, or for $\mathcal{R}$ the exterior of a Schwarzschild black hole and for the Hartle--Hawking state~\cite{sewell1982,kay1985a,kay1985b,kaywald1991}. For conformal theories, local geometric flows are also obtained for light cone and diamond (or double cone) regions, both in Minkowski~\cite{buchholz1977,hisloplongo1982,hislop1988} and general conformally flat spacetimes~\cite{casinihuertamyers2011,froeb2023}. Many more modular Hamiltonians are known for one- and two-dimensional quantum field theories in a variety of settings, and we refer the reader to Ref.~\cite{froeb2023} for a (possibly incomplete) list. However, these cases are rather the exception and in general the flow can be highly non-local. In fact, it has been argued that a local geometric modular flow~\cite{buchholzsummers1993,buchholzdreyerflorigsummers2000} can only be a (conformal) symmetry of the background spacetime~\cite{sorce2024}.

It is not surprising that non-local modular flows are harder to determine, and explicit examples are only known in one and two dimensions. In particular, for free massless fermions in the Minkowski vacuum state and with $\mathcal{R}$ given by the union of a collection of diamonds, the flow becomes multi-local and couples fermions of different diamonds~\cite{casinihuerta2009b,rehrentedesco2013}. The same happens for general chiral conformal field theories on the light ray~\cite{longomartinettirehren2010,hollands2021}, while for a state of finite temperature (including the zero-temperature limit) on a cylinder the modular Hamiltonian and the modular flow become highly non-local~\cite{klichvamanwong2017,blancopereznadal2019,friesreyes2019,blancogarbarzpereznadal2019}. For massive fermions, it seems only possible to obtain numerical~\cite{bostelmanncadamurominz2023} or perturbative results~\cite{ariasblancocasinihuerta2017,cadamurofroebminz2023}. These again show complete non-locality, apart from the leading singular terms which appear to be universal and in fact agree with the massless result. The explicit results for free fermions were obtained using a formula relating the one-particle modular Hamiltonian to the two-point function, and which can be derived in a variety of ways~\cite{araki1970,peschel2003,casinihuerta2009a,hollands2021}. This formula reads
\begin{equation}
\label{eq:formula_hg}
H = \ln\left( \frac{G}{1-G} \right) \eqend{,}
\end{equation}
where $H$ is the integral kernel of the modular Hamiltonian $\ln \Delta_\Omega$ and $G$ the two-point function of the state $\Omega$ restricted to $\mathcal{R}$, and we will also employ it for our results.

Concretely, we consider the modular flow for massless Dirac fermions in 1+1 dimensions on a cylinder. This system has been studied previously for antiperiodic (Neveu--Schwarz) boundary conditions~\cite{rehrentedesco2013}, where the ground state is unique and the modular flow is local. It has also been studied for periodic (Ramond) boundary conditions~\cite{klichvamanwong2017,blancopereznadal2019,friesreyes2019,blancogarbarzpereznadal2019} and the zero-temperature limit of a thermal state, where the modular flow is completely non-local. In both cases, the flow does not mix the two chiralities of the fermion. However, for periodic boundary conditions there is no unique ground state, and the zero-energy state is highly degenerate. The modular flow and modular Hamiltonian for a generic quasi-free (Gaussian) zero-energy state show very interesting behavior, which we determine here for the first time. Namely, we prove the following:
\begin{theorem}
\label{thm:twopf}
Consider a free massless Dirac fermion $\psi$ on the cylinder $\mathbb{S}^1 \times \mathbb{R}$ of circumference $L$, with the single-particle Hilbert space $\mathcal{H} = L^2([0,L]) \oplus L^2([0,L])$ of initial data at time $t = 0$ and the corresponding Fock space. For antiperiodic boundary conditions $\psi(x+L) = - \psi(x)$, there exists a unique quasi-free ground state $\omega_\text{NS}$ whose two-point function at $t = 0$ reads
\begin{equation*}
\omega_\text{NS}\left( \psi(f) \left[ \psi(g) \right]^\dagger \right) = \frac{1}{2 \mathi L} \lim_{\epsilon \to 0^+} \iint_0^L \left[ \frac{f_1(x) g^*_1(y)}{\sin\left[ \frac{\pi}{L} (x-y-\mathi \epsilon) \right]} - \frac{f_2(x) g^*_2(y)}{\sin\left[ \frac{\pi}{L} (x-y+\mathi \epsilon) \right]} \right] \total x \total y \eqend{,}
\end{equation*}
where $f_a, g_a \in L^2([0,L])$ with $a = 1,2$ are test functions for each chirality. For periodic boundary conditions $\psi(x+L) = \psi(x)$, there exists a four-parameter family of quasi-free ground states $\omega_{\text{R},h}$ whose two-point function at $t = 0$ reads
\begin{equation*}
\begin{split}
\omega_{\text{R},h}\left( \psi(f) \left[ \psi(g) \right]^\dagger \right) &= \frac{1}{2 \mathi L} \lim_{\epsilon \to 0^+} \iint_0^L \left[ \frac{f_1(x) g^*_1(y)}{\tan\left[ \frac{\pi}{L} (x-y-\mathi \epsilon) \right]} - \frac{f_2(x) g^*_2(y)}{\tan\left[ \frac{\pi}{L} (x-y+\mathi \epsilon) \right]} \right] \total x \total y \\
&\quad+ \sum_{a,b=1}^2 h_{ab} \int_0^L f_a(x) \total x \int_0^L g^*_b(y) \total y \eqend{,}
\end{split}
\end{equation*}
and where the matrix $h$ is given by
\begin{equation*}
h = \frac{h_1 + h_2}{2} \1 + \frac{h_1 - h_2}{2} \begin{pmatrix} \cos \psi & \sin \psi \, \mathe^{\mathi \phi} \\ \sin \psi \, \mathe^{- \mathi \phi} & - \cos \psi \end{pmatrix}
\end{equation*}
with $\abs{h_i} \leq \frac{1}{2 L}$ for $i \in \{1,2\}$, $\phi \in [0,2\pi)$ and $\psi \in [0,\pi]$. The state is pure if $\abs{h_1} = \abs{h_2} = \frac{1}{2 L}$ and mixed otherwise. The zero-temperature limit of the thermal state studied in~\cite{klichvamanwong2017,blancopereznadal2019,friesreyes2019,blancogarbarzpereznadal2019} corresponds to $h_1 = h_2 = 0$, while the massless limit of the vacuum state for massive Dirac fermions results in $\phi = \psi = \frac{\pi}{2}$ and $h_2 = - h_1 = \frac{1}{2 L}$. All states $\omega_{\text{R},h}$ have zero energy. They lie inside a double cone whose tips correspond to the pure states with $h_1 = h_2 = \pm \frac{1}{2 L}$, and whose rim corresponds to the pure states with $h_1 = - h_2 = \pm \frac{1}{2 L}$.
\end{theorem}
The freedom in the choice of state that is parametrized by $h$ corresponds to the zero mode of the fermion, which is absent for antiperiodic boundary conditions. Employing the formula~\eqref{eq:formula_hg} and its generalization for the modular flow, we then prove:
\begin{theorem}
\label{thm:flow}
In the setting of Theorem~\ref{thm:twopf}, we consider the von Neumann algebra $\mathfrak{A}(\mathcal{R})$ generated by fermions $\psi(f)$ with $\supp f \subset \mathcal{R}$. The modular operator $\Delta_{\omega,\mathcal{R}}$ associated to this algebra and a quasi-free state $\omega$ induces a modular flow $\sigma_t(A) = \Delta_{\omega,\mathcal{R}}^{\mathi t} A \Delta_{\omega,\mathcal{R}}^{- \mathi t}$ for $A \in \mathfrak{A}(\mathcal{R})$. On the dense subspace $\mathfrak{D}(\mathcal{R}) \subset \mathfrak{A}(\mathcal{R})$ spanned by products of a finite number of fermion operators smeared with Schwartz functions $f_i \in \mathcal{S}(\mathcal{R})$, its action is given by $\sigma_t\left( \psi(f_1) \cdots \psi(f_n) \right) = \sigma_t\left( \psi(f_1) \right) \cdots \sigma_t\left( \psi(f_n) \right)$, where
\begin{equation}
\label{eq:flow_singleparticle}
\sigma_t\left( \psi(f) \right) = \psi(f(t)) \eqend{,} \quad f_a(t,x) \equiv \sum_{b=1}^2 \int_{\mathcal{R}} K_{ab}(t,x,y) f_b(y) \total y \quad (a=1,2)
\end{equation}
with a kernel $K$ that depends on the state $\omega$ and the region $\mathcal{R}$. For $\mathcal{R} = [-\ell,\ell]$ an interval of length $2 \ell < L$ and the antiperiodic ground state $\omega_\text{NS}$, we have
\begin{equation}
\label{eq:flow_kernel_ns}
K^\text{NS}_{ab}(t,x,y) = \frac{2 \pi}{L} \frac{\sinh(\pi t)}{\sin\left[ \frac{\pi}{L} (x-y) \right]} \delta\big[ 2 \pi t - \Omega_a(x) + \Omega_a(y) \big] \begin{pmatrix} 1 & 0 \\ 0 & - 1 \end{pmatrix}_{ab}
\end{equation}
with
\begin{equation}
\label{eq:flow_omega_def}
\Omega_1(x) \equiv \ln\left( \frac{\sin\left[ \frac{\pi}{L} (\ell+x) \right]}{\sin\left[ \frac{\pi}{L} (\ell-x) \right]} \right) \eqend{,} \quad \Omega_2(x) \equiv - \Omega_1(x) = \Omega_1(-x) \eqend{,}
\end{equation}
and the flow is local. For the same region $\mathcal{R} = [-\ell,\ell]$ and the family of states $\omega_{\text{R},h}$, we have instead
\begin{splitequation}
\label{eq:flow_kernel_r}
K^{\text{R},h}_{ab}(t,x,y) &= K^\text{NS}_{ab}(t,x,y) + \frac{1}{4 \ell} \sinh(\pi t) \, \pf \frac{1}{\sinh\left[ \frac{L}{4 \ell} \big[ 2 \pi t - \Omega_a(x) + \Omega_b(y) \big] \right]} \\
&\qquad\times \Bigg[ \left( \frac{1 + 2 L h_1}{1 - 2 L h_1} \right)^\frac{\mathi \big[ 2 \pi t - \Omega_a(x) + \Omega_b(y) \big] L}{4 \pi \ell} \begin{pmatrix} 1 + \cos \psi & \sin \psi \, \mathe^{\mathi \phi} \\ \sin \psi \, \mathe^{- \mathi \phi} & 1 - \cos \psi \end{pmatrix}_{ab} \\
&\qquad\quad+ \left( \frac{1 + 2 L h_2}{1 - 2 L h_2} \right)^\frac{\mathi \big[ 2 \pi t - \Omega_a(x) + \Omega_b(y) \big] L}{4 \pi \ell} \begin{pmatrix} 1 - \cos \psi & - \sin \psi \, \mathe^{\mathi \phi} \\ - \sin \psi \, \mathe^{- \mathi \phi} & 1 + \cos \psi \end{pmatrix}_{ab} \Bigg] \eqend{,}
\end{splitequation}
and the flow has a local and a non-local part. On the full algebra $\mathfrak{A}(\mathcal{R})$ the modular flow $\sigma_t$ is then defined by continuity.
\end{theorem}
While we restrict to an interval for the sake of simplicity, the generalization to an arbitrary set of disjoint intervals is straightforward. As long as $\abs{h_i} \neq \frac{1}{2L}$ for $i \in \{1,2\}$, which is the case for a mixed state, the flow is clearly non-local. In the limit $h_i \to \pm \frac{1}{2 L}$ where the state becomes pure, the factors $[ (1 + 2 L h_i)/(1 - 2 L h_i) ]^{\mathi \alpha}$ oscillate rapidly, and one expects the character of the flow to change. To determine the limit, we employ the following lemma, which could be of independent interest:
\begin{lemma}
\label{lemma:limit}
The following limits exist in the weak topology of tempered distributions, i.e., after smearing with a Schwartz function $f \in \mathcal{S}(\mathbb{R})$:
\begin{equation*}
\lim_{a \to 0} \left[ a^{\mathi t} \pf \frac{1}{\sinh(\pi t)} - \mathi \frac{a-1}{a+1} \delta(t) \right] = \lim_{a \to \infty} \left[ a^{\mathi t} \pf \frac{1}{\sinh(\pi t)} - \mathi \frac{a-1}{a+1} \delta(t) \right] = 0 \eqend{.}
\end{equation*}
The limits also hold for all continuous $f$ satisfying $\int_{-\infty}^\infty \abs{ \frac{f(t)}{\sinh(\pi t)} } \total t < \infty$.
\end{lemma}
From the results of Theorem~\ref{thm:flow} we then obtain the following limit:
\begin{corollary}
\label{cor:flow}
In the setting of Theorem~\ref{thm:flow}, considering the region $\mathcal{R} = [-\ell,\ell]$ and the family of states $\omega_{\text{R},h}$, the following limits exist in the weak topology:
\begin{splitequation}
\lim_{h_1 = h_2 \to \pm \frac{1}{2 L}} K^{\text{R},h}_{ab}(t,x,y) &= \frac{2 \pi}{L} \frac{\sinh(\pi t)}{\sin\left[ \frac{\pi}{L} (x-y) \right]} \delta\big[ 2 \pi t - \Omega_a(x) + \Omega_b(y) \big] \begin{pmatrix} 1 & 0 \\ 0 & - 1 \end{pmatrix}_{ab} \\
&\quad\pm \frac{2 \pi \mathi}{L} \sinh(\pi t) \, \delta\big[ 2 \pi t - \Omega_a(x) + \Omega_b(y) \big] \delta_{ab} \eqend{,}
\end{splitequation}
which corresponds to the pure states at the tip of the double cone, and
\begin{splitequation}
\lim_{h_1 = - h_2 \to \pm \frac{1}{2 L}} K^{\text{R},h}_{ab}(t,x,y) &= \frac{2 \pi}{L} \frac{\sinh(\pi t)}{\sin\left[ \frac{\pi}{L} (x-y) \right]} \delta\big[ 2 \pi t - \Omega_a(x) + \Omega_b(y) \big] \begin{pmatrix} 1 & 0 \\ 0 & - 1 \end{pmatrix}_{ab} \\
&\quad\pm \frac{2 \pi \mathi}{L} \sinh(\pi t) \, \delta\big[ 2 \pi t - \Omega_a(x) + \Omega_b(y) \big] \begin{pmatrix} \cos \psi & \sin \psi \, \mathe^{\mathi \phi} \\ \sin \psi \, \mathe^{- \mathi \phi} & - \cos \psi \end{pmatrix}_{ab} \eqend{,}
\end{splitequation}
which corresponds to the pure states at the rim.
\end{corollary}
We see that for pure states, the modular flow is again local, but that it has a different form in comparison to the antiperiodic case.

To obtain the modular Hamiltonian $H = \ln \Delta$, we compute the generator of the flow. With the explicit expressions above, this is almost trivial, except for the fact that in the local terms, as $t \to 0$ the flow moves $y$ to $x$ and the zero of $\sinh(\pi t)$ cancels the singularity of $\frac{1}{\sin\left[ \frac{\pi}{L} (x-y) \right]}$ to leave a finite result. We prove:
\begin{theorem}
\label{thm:hamiltonian}
Consider the setting of Theorem~\ref{thm:flow}. The modular Hamiltonian $\ln \Delta_{\omega,\mathcal{R}}$, the generator of the modular flow $\sigma_t$, is defined on the dense subspace of Fock space obtained by acting with elements of $\mathfrak{D}(\mathcal{R})$ on the ground state. It acts there by annihilating the ground state, and on other states via the commutator
\begin{equation*}
[ \ln \Delta_{\omega,\mathcal{R}}, \psi(f) ] = \psi(\hat{f}) \eqend{,} \quad \hat{f}_a(t,x) \equiv \sum_{b=1}^2 \int_{\mathcal{R}} H_{ab}(x,y) f_b(y) \total y \quad (a=1,2)
\end{equation*}
with a kernel $H$ that depends on the ground state and the region $\mathcal{R}$. For $\mathcal{R} = [-\ell,\ell]$ an interval of length $2 \ell < L$ and the antiperiodic ground state $\ket{0_\text{NS}}$, we have
\begin{equation}
\label{eq:hamiltonian_kernel_ns}
H^\text{NS}_{ab}(x,y) = 2 \mathi L \frac{\sin^2\left( \frac{\pi}{L} \ell \right) - \sin\left( \frac{\pi}{L} x \right) \sin\left( \frac{\pi}{L} y \right)}{\sin\left( \frac{2 \pi \ell}{L} \right)} \delta'(x-y) \begin{pmatrix} 1 & 0 \\ 0 & - 1 \end{pmatrix}_{ab} \eqend{,}
\end{equation}
which is local. For the same region $\mathcal{R} = [-\ell,\ell]$ and the family of states $\omega_{\text{R},h}$, we have instead
\begin{splitequation}
\label{eq:hamiltonian_kernel_r}
H^{\text{R},h}_{ab}(x,y) &= H^\text{NS}_{ab}(x,y) + \frac{\mathi \pi}{4 \ell} \, \pf \frac{1}{\sinh\left[ \frac{L}{4 \ell} \big[ \Omega_a(x) - \Omega_b(y) \big] \right]} \\
&\qquad\times \Bigg[ \left( \frac{1 + 2 L h_1}{1 - 2 L h_1} \right)^\frac{-\mathi \big[ \Omega_a(x) - \Omega_b(y) \big] L}{4 \pi \ell} \begin{pmatrix} 1 + \cos \psi & \sin \psi \, \mathe^{\mathi \phi} \\ \sin \psi \, \mathe^{- \mathi \phi} & 1 - \cos \psi \end{pmatrix}_{ab} \\
&\qquad\quad+ \left( \frac{1 + 2 L h_2}{1 - 2 L h_2} \right)^\frac{- \mathi \big[ \Omega_a(x) - \Omega_b(y) \big] L}{4 \pi \ell} \begin{pmatrix} 1 - \cos \psi & - \sin \psi \, \mathe^{\mathi \phi} \\ - \sin \psi \, \mathe^{- \mathi \phi} & 1 + \cos \psi \end{pmatrix}_{ab} \Bigg] \eqend{,}
\end{splitequation}
whose local part is the same as before, but which has an additional non-local part. As usual, the domain of definition of the modular Hamiltonian can be extended to all vectors in Fock space obtained by acting with $A \in \mathfrak{A}(\mathcal{R})$ on the ground state for which the limit
\begin{equation*}
\lim_{t \to 0^+} \frac{1}{\mathi t} \big[ \sigma_t(A) - A \big] \ket{0}
\end{equation*}
exists. This includes in particular the case where $A$ is a linear combination of products of fermion operators smeared with functions $f_i \in L^2([-\ell,\ell]) \oplus L^2([-\ell,\ell])$ for which also $\hat{f}_i \in L^2([-\ell,\ell]) \oplus L^2([-\ell,\ell])$.
\end{theorem}
In the cases that have been studied before, namely antiperiodic boundary conditions and the zero-temperature limit of the thermal state for periodic boundary conditions (which has $h_1 = h_2 = 0$), our results agree with the known ones. For periodic boundary conditions and the general zero-energy state $\omega_{\text{R},h}$, our results are completely new. Both the modular flow and the modular Hamiltonian change character if we pass from a mixed to a pure state. For a mixed state we have complete non-locality, i.e., no matter how strongly localized the test function $f$ is that the fermion operator is smeared with, the commutator with the modular Hamiltonian results in a fermion operator smeared with a test function $\hat{f}$ that has support in the full interval $[-\ell,\ell]$. For a pure state on the other hand, we again can use Lemma~\ref{lemma:limit} to obtain
\begin{corollary}
\label{cor:hamiltonian}
In the setting of Theorem~\ref{thm:hamiltonian}, considering the region $\mathcal{R} = [-\ell,\ell]$ and the family of states $\omega_{\text{R},h}$, the following limits exist in the weak topology:
\begin{splitequation}
\label{eq:hamiltonian_limit_eq}
\lim_{h_1 = h_2 \to \pm \frac{1}{2 L}} H^{\text{R},h}_{ab}(x,y) &= H^\text{NS}_{ab}(x,y) \pm 2 \pi \frac{\sin^2\left( \frac{\pi}{L} \ell \right) - \sin^2\left( \frac{\pi}{L} x \right)}{\sin\left( \frac{2 \pi \ell}{L} \right)} \delta(x-y) \delta_{ab} \eqend{,}
\end{splitequation}
which corresponds to the pure states at the tip of the double cone, and
\begin{splitequation}
\label{eq:hamiltonian_limit_neq}
\lim_{h_1 = - h_2 \to \pm \frac{1}{2 L}} H^{\text{R},h}_{ab}(x,y) &= H^\text{NS}_{ab}(x,y) \pm 2 \pi \frac{\sin^2\left( \frac{\pi}{L} \ell \right) - \sin^2\left( \frac{\pi}{L} x \right)}{\sin\left( \frac{2 \pi \ell}{L} \right)} \\
&\qquad\times \left[ \cos \psi \, \delta(x-y) \begin{pmatrix} 1 & 0 \\ 0 & - 1 \end{pmatrix}_{ab} + \sin \psi \, \delta(x+y) \begin{pmatrix} 0 & \mathe^{\mathi \phi} \\ \mathe^{- \mathi \phi} & 0 \end{pmatrix}_{ab} \right] \eqend{,}
\end{splitequation}
which corresponds to the pure states at the rim.
\end{corollary}
For pure states, the modular Hamiltonian is thus local or (for $\sin \psi \neq 0$) anti-local, i.e., it transforms fermion operators smeared with a function localized around some point $x$ into a sum of fermion operators smeared with functions localized around $x$ and $-x$. This is an effect that is also seen numerically for massive fermions~\cite{bostelmanncadamurominz2024}, and in fact the massless limit of the vacuum state for massive Dirac fermions corresponds to the state with $\phi = \psi = \frac{\pi}{2}$ and $h_2 = - h_1 = \frac{1}{2 L}$.

In the remainder of this work, we will give the proofs of the above theorems. Theorem~\ref{thm:twopf} is proven in section~\ref{sec:prop}, using standard methods of canonical quantization. To prove Theorem~\ref{thm:flow}, we first note that for free theories, the modular flow and modular Hamiltonian are second-quantized operators on Fock space~\cite{eckmannosterwalder1973,figlioliniguido1989} and it is enough to determine the single-particle operators. From the formula~\eqref{eq:formula_hg} which gives the integral kernel of the single-particle modular Hamiltonian, it follows that
\begin{equation}
\label{eq:formula_kg}
K = \mathe^{\mathi t H} = \left( \frac{G}{1-G} \right)^{\mathi t}
\end{equation}
is the integral kernel of the modular flow for single-particle states. The right-hand side can be defined via spectral theory, which presents no difficulties since $G$ [as a convolution operator on $L^2([-\ell,\ell]) \oplus L^2([-\ell,\ell])$] is bounded with absolutely continuous spectrum contained in $[0,1]$, and $0$ and $1$ are not eigenvalues~\cite[Corollary~4.10]{araki1970}. To determine the spectral measure of $G$, we use Stone's formula~\cite[Theorem~VII.13]{reedsimon1} for an absolutely continuous spectrum
\begin{equation}
\label{eq:stone_formula}
\total E_A(\mu) = \frac{1}{2 \pi \mathi} \lim_{\epsilon \to 0^+} \Big[ R_A(\mu + \mathi \epsilon) - R_A(\mu - \mathi \epsilon) \Big] \total \mu \eqend{,}
\end{equation}
where $E_A$ is the spectral measure of $A$ supported on the spectrum $\sigma(A)$, where $R_A(\mu) \equiv (A - \mu \1)^{-1}$ is the resolvent of $A$, and where the limit is taken in the strong operator topology (i.e., acting on vectors in Hilbert space). We thus obtain that the function $f_a(t,x)$ in~\eqref{eq:flow_singleparticle} is given by
\begin{equation}
\label{eq:flow_singleparticle_function}
f_a(t,x) = \frac{1}{2 \pi \mathi} \int_0^1 \left( \frac{\mu}{1-\mu} \right)^{\mathi t} \lim_{\epsilon \to 0^+} \sum_{b=1}^2 \int_{-\ell}^\ell \Big[ R(\mu + \mathi \epsilon)_{ab}(x,y) - R(\mu - \mathi \epsilon)_{ab}(x,y) \Big] f_b(y) \total y \total \mu \eqend{,}
\end{equation}
where now $R(\mu)_{ab}(x,y)$ is the integral kernel of the resolvent of the two-point function $G$, again seen as a convolution operator that acts on the space $L^2([-\ell,\ell]) \oplus L^2([-\ell,\ell])$ of initial data at $t = 0$. We therefore have to determine the resolvent, which we do in section~\ref{sec:resolvent} for the two-point functions in the different states $\omega_\text{NS}$ and $\omega_{\text{R},h}$. For this, we leverage Carleman's method~\cite{carleman1922} (see also the review~\cite{estradakanwal1987}), which consists in mapping the resolvent equation to a Riemann--Hilbert problem. This approach was also used to compute the modular Hamiltonian for finite-temperature states on the cylinder~\cite{klichvamanwong2017,blancopereznadal2019,friesreyes2019,blancogarbarzpereznadal2019}, and for the spectral decomposition of the finite Hilbert transform~\cite{koppelmanpincus1959}, which is relevant for diamond regions in Minkowski~\cite{casinihuerta2009a,casinihuerta2009b,rehrentedesco2013} and chiral CFTs on the light ray~\cite{longomartinettirehren2010,hollands2021}. The proof of Theorem~\ref{thm:flow} is then finished in section~\ref{sec:flow}, where we explicitly compute the function $f_a(t,x)$~\eqref{eq:flow_singleparticle_function} and extract the kernel $K_{ab}(t,x,y)$. The proof of Lemma~\ref{lemma:limit} is relegated to appendix~\ref{app:lemma_limit}, while the easy proofs of Corollary~\ref{cor:flow}, Theorem~\ref{thm:hamiltonian} and Corollary~\ref{cor:hamiltonian} are given at the end of section~\ref{sec:flow}. We close with an outlook in section~\ref{sec:outlook}.

\section{Boundary conditions, states and propagators}
\label{sec:prop}

In this section, we are giving the proof of Theorem~\ref{thm:twopf}. As explained in the introduction, we are considering a free massless Dirac field $\psi$ in 1+1 dimensions on the cylinder $\mathcal{C}_L \equiv \mathbb{S}^1 \times \mathbb{R}$, where the spatial circle has circumference $L$. We identify the cylinder with the compactified plane $\mathbb{R}^2 / \sim$, where two points are equivalent if their spatial coordinates differ by a multiple of the circumference: $(x,t) \sim (x',t') \ \Leftrightarrow\  t = t', x = x' + k L$ with $k \in \mathbb{Z}$. The fermion field has two complex components $\psi_a$, $a \in \{1,2\}$, and the equation of motion implies that each component is a function of a single null coordinate: $\psi_1(x,t) = f(x+t)$ and $\psi_2(x,t) = g(x-t)$. Since fermion fields are not observable but quantities bilinear in the field (such as the current) are, we can choose either periodic or antiperiodic boundary conditions $\psi(x+L,t) = \pm \psi(x,t)$. The first (periodic) choice is usually called Ramond (R) boundary condition, while the second (antiperiodic) one is known as Neveu--Schwarz (NS) boundary condition. As the single-particle Hilbert space we take $\mathcal{H} = L^2([0,L]) \oplus L^2([0,L])$, the space of initial data at time $t = 0$ with the appropriate boundary condition, and then the usual fermionic Fock space $\mathcal{F}$ constructed from it.

Since the $t = 0$ Cauchy surface is compact, individual modes of the fermion field are well-defined operators on Fock space. For periodic boundary conditions, these are defined as
\begin{equation}
\psi_{an} \equiv \frac{1}{\sqrt{L}} \int_0^L \psi_a(x,0) \, \mathe^{- \mathi \frac{2 \pi n}{L} x} \total x \eqend{,}
\end{equation}
with $n \in \mathbb{Z}$, while for antiperiodic boundary conditions we have to take $n + \frac{1}{2} \in \mathbb{Z}$. The spatially smeared fermion field at time $t = 0$ can then be expanded as
\begin{equation}
\label{eq:fourier_r}
\psi(f) = \sum_{a=1}^2 \sum_{n \in \mathbb{Z}} \tilde{f}_a(n) \, \psi_{an}
\end{equation}
in the periodic case, or
\begin{equation}
\label{eq:fourier_ns}
\psi(f) = \sum_{a=1}^2 \sum_{n \in \mathbb{Z}} \tilde{f}_a\left( n + \frac{1}{2} \right) \, \psi_{a,n+\frac{1}{2}}
\end{equation}
in the antiperiodic case, where we defined the Fourier transform
\begin{equation}
\label{eq:fourier_def}
\tilde{f}_a(n) \equiv \frac{1}{\sqrt{L}} \int_0^L f_a(x) \, \mathe^{\mathi \frac{2 \pi n}{L} x} \total x \eqend{.}
\end{equation}
The canonical anticommutation relations
\begin{equation}
\label{eq:car}
\{ \psi(f), [ \psi(g) ]^\dagger \} = \sum_a \int_0^L f_a(x) g_a^*(x) \total x \eqend{,} \quad \{ \psi(f), \psi(g) \} = 0
\end{equation}
imply the anticommutators $\{ \psi_{an}^{\vphantom{\dagger}}, \psi^\dagger_{bm} \} = \delta_{ab} \delta_{mn}$ and $\{ \psi_{an}, \psi_{bm} \} = 0$ for the modes in both the periodic and antiperiodic cases. The Hamiltonian is given by the quadratic form
\begin{equation}
\label{eq:hamiltonian}
E \equiv \frac{2 \pi}{L} \sum_{n \neq 0} \abs{n} \left( a_n^\dagger a_n^{\vphantom{\dagger}} + b_n^\dagger b_n^{\vphantom{\dagger}} \right) \eqend{,}
\end{equation}
where
\begin{equation}
\label{eq:annihilation}
a_n \equiv \begin{cases} \psi_{1,n} & n < 0 \\ \frac{1}{\sqrt{2}} \left( \psi_{10} - \mathi \psi_{20} \right) & n = 0 \\ \psi_{2,n} & n > 0 \eqend{,} \end{cases} \qquad b_n \equiv \begin{cases} \psi_{1,-n}^\dagger & n < 0 \\ \frac{1}{\sqrt{2}} \left( \psi_{10}^\dagger - \mathi \psi_{20}^\dagger \right) & n = 0 \\ \psi_{2,-n}^\dagger & n > 0 \eqend{,} \end{cases}
\end{equation}
and it is well-defined (at least) on the usual subspace of Fock space whose vectors only have a finite number of non-zero entries. From the anticommutators of the modes, we obtain the anticommutators
\begin{equation}
\{ a_n^{\vphantom{\dagger}}, a_m^\dagger \} = \delta_{nm} = \{ b_n^{\vphantom{\dagger}}, b_m^\dagger \} \eqend{,}
\end{equation}
with all other anticommutators between the $a_n$ and $b_m$ vanishing. Using the expansions~\eqref{eq:fourier_r}, \eqref{eq:fourier_ns}, it is then easy to verify that $[ E, \psi(f) ] = \psi(f_E)$ with $(f_E)_1(x) = \mathi f'_1(x)$ and $(f_E)_2(x) = - \mathi f'_2(x)$ for both periodic and antiperiodic boundary conditions. Since $\psi_1(x,t)$ only depends on $x+t$ and $\psi_2(x,t)$ only depends on $x-t$, this is equivalent to the Heisenberg equations of motion $[ E, \psi ] = - \mathi \partial_t \psi$ for the unsmeared fermion field. For later use, we note that the inverse relations to~\eqref{eq:annihilation} are given by
\begin{equation}
\label{eq:psi_in_annihilation_creation}
\psi_{1,n} = \begin{cases} a_n & n < 0 \\ \frac{1}{\sqrt{2}} \left( a_0 + b_0^\dagger \right) & n = 0 \\ b_{-n}^\dagger & n > 0 \eqend{,} \end{cases} \qquad \psi_{2,n} = \begin{cases} b_{-n}^\dagger & n < 0 \\ \frac{\mathi}{\sqrt{2}} \left( a_0 - b_0^\dagger \right) & n = 0 \\ a_n & n > 0 \eqend{.} \end{cases}
\end{equation}

Note that while all modes appear in the expansion of the Hamiltonian~\eqref{eq:hamiltonian} in the antiperiodic case, for periodic boundary conditions the modes with $n = 0$ are missing. It follows that in the antiperiodic (Neveu--Schwarz) case, there is a unique ground state $\ket{0_\text{NS}} \in \mathcal{F}$ which is annihilated by all the $a_n$ and $b_n$, and consequently has zero energy. On the other hand, in the periodic (Ramond) case, the ground state is four-fold degenerate: the states of zero energy must be annihilated by the $a_n$ and $b_n$ for $n > 0$, but the occupation number of the zero modes $\psi_{a,0}$ can be arbitrary. Among all the possible ground states, the vector state that is obtained by starting with the vacuum of the massive theory (which is unique) and then sending the mass to zero is the one that is also annihilated by $a_0$ and $b_0$, and we will refer to it as the \emph{massive vacuum} $\ket{0_\text{m}} \in \mathcal{F}$. In the massive theory, the three zero-momentum states $a_0^\dagger \ket{0_\text{m}}$, $b_0^\dagger \ket{0_\text{m}}$ and $a_0^\dagger b_0^\dagger \ket{0_\text{m}}$ with zero momentum have energy proportional to the mass, so they also become ground states when the mass is sent to zero. Together with the massive vacuum, they form an orthonormal basis of the \emph{zero-mode sector}, the eigenspace of $E$ corresponding to the eigenvalue 0. Another way to pick a ground state is to start with a thermal state and then send the temperature to zero. This yields a maximally mixed state, where each of the zero-momentum states and the massive vacuum occur with probability $\frac{1}{4}$. We will refer to this state as the \emph{zero-temperature state}. This is the state that was considered in previous work~\cite{klichvamanwong2017,blancopereznadal2019,friesreyes2019,blancogarbarzpereznadal2019}.

\begin{figure}[ht]
\centering
\includegraphics[scale=0.6]{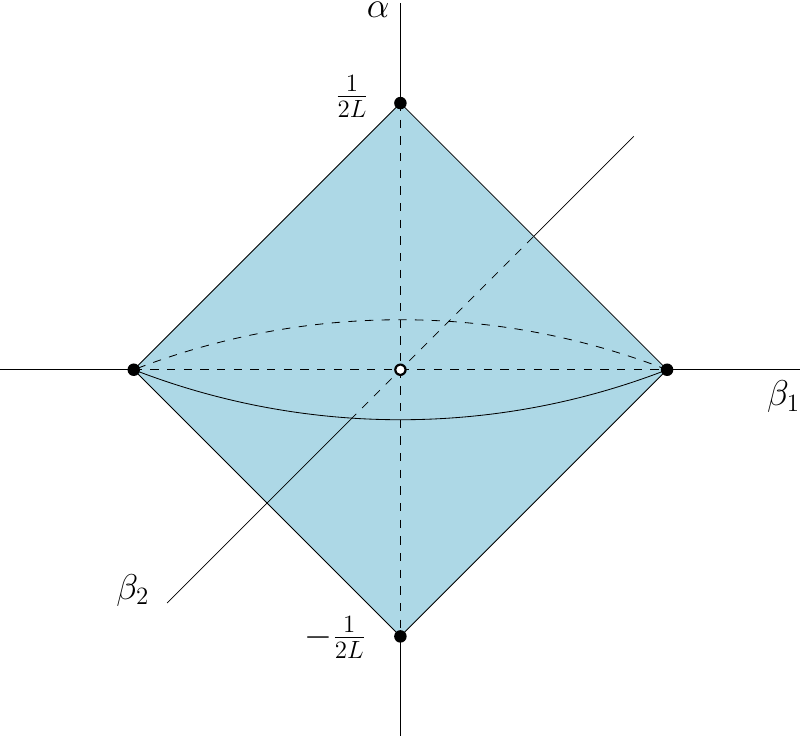}
\caption{The set of all quasi-free (Gaussian) ground states for periodic boundary conditions can be visualized as a double cone in $\mathbb{R}^4$, which we represent here with one dimension suppressed. The extreme points of this convex set (the two tips of the cones and the rim where both cones meet) correspond to pure states, and the remaining points represent mixed states.}
\label{fig:states}
\end{figure}

In this work, we consider a generic mixture of ground states for periodic boundary conditions. Since the zero-mode sector is finite-dimensional, we can represent this state as a vector state $\ket{0_\text{R}}$ on the subspace of Fock space where the zero modes are unoccupied, and a density matrix $\rho_{\text{R},h}$ in the zero-mode sector.\footnote{To apply the results of Tomita--Takesaki theory, in principle one would have to purify the state on a doubled Hilbert space for the zero-mode sector and work with a vector state $\ket{0_{\text{R},h}}$ on the corresponding Fock space. Since this is a standard procedure and does not influence the derivation of our results which only depends on the two-point function according to Eqs.~\eqref{eq:formula_hg} and~\eqref{eq:formula_kg}, we leave the details to the reader.} The resulting state satisfies
\begin{equation}
\label{eq:ground}
a_n \ket{0_\text{R}} = b_n \ket{0_\text{R}} = 0 \quad\text{for all}\quad n \neq 0 \eqend{,}
\end{equation}
and includes the states discussed above as particular cases. Besides~\eqref{eq:ground}, we will impose that the state be quasi-free (or Gaussian), such that it is completely characterized by the covariance matrix of the zero-mode sector, which is the $2 \times 2$ matrix $g$ with components
\begin{equation}
\label{eq:matrix_g}
g_{ab} \equiv \frac{1}{L} \omega_{\text{R},h}\left( \psi_{a0}^{\vphantom{\dagger}} \, \psi_{b0}^\dagger \right) = \frac{1}{L} \tr\left( \rho_{\text{R},h} \, \psi_{a0}^{\vphantom{\dagger}} \, \psi_{b0}^\dagger \right) \eqend{.}
\end{equation}
The zero-mode sector can be obtained from the full smeared fermion field $\psi(f)$ by choosing $f_a(x) = \text{const}$ in Eqs.~\eqref{eq:fourier_def} and~\eqref{eq:fourier_r}, and the zero modes are seen to decouple from the other modes and to fulfill separately the canonical anticommutation relations. The positivity of the state $\omega_{\text{R},h}$ together with the anticommutation relations then imply that $L g$ is a Hermitean matrix with eigenvalues between 0 and 1~\cite[Lemma~3.2]{araki1970}, and it follows that the matrix $h \equiv g - \frac{1}{2 L} \1$ is Hermitean with eigenvalues between $- \frac{1}{2 L}$ and $\frac{1}{2 L}$. Any Hermitean $2 \times 2$ matrix can be written as a real-linear combination of the identity and the three Pauli matrices, such that
\begin{equation}
\label{eq:h_pauli}
h = \alpha \1 + \vec{\beta} \cdot \vec{\sigma}
\end{equation}
with $\alpha \in \mathbb{R}$ and $\vec{\beta} \in \mathbb{R}^3$. The eigenvalues of $h$ are given by $\alpha \pm \abs{\vec{\beta}}$, such that we obtain the condition
\begin{equation}
\label{eq:condstates}
\abs{\alpha} + \abs{\vec{\beta}} \leq \frac{1}{2 L} \eqend{.}
\end{equation}
It follows that the set of allowed values of $h$ (or, in other words, the set of states we are considering) can be visualized as a double cone, se Fig.~\ref{fig:states}.

Since for a $2 \times 2$ matrix we have $\tr h = h_1 + h_2$ and $\det h = h_1 h_2$ with the two eigenvalues $h_i$, $i \in \{1,2\}$ of $h$, it follows that $\alpha = \frac{h_1 + h_2}{2}$ and $\abs{\vec{\beta}} = \frac{\abs{h_1 - h_2}}{2}$. Using then spherical coordinates for $\vec{\beta}$, we obtain
\begin{equation}
\label{eq:h_decomp}
h = \frac{h_1 + h_2}{2} \1 + \frac{h_1 - h_2}{2} \begin{pmatrix} \cos \psi & \sin \psi \, \mathe^{\mathi \phi} \\ \sin \psi \, \mathe^{- \mathi \phi} & - \cos \psi \end{pmatrix}
\end{equation}
with $\phi \in [0,2\pi)$ and $\psi \in [0,\pi]$, and where we absorbed the sign of $h_1 - h_2$ in $\psi$. The extreme points of this convex set, i.e., the tips of the cones and the rim where both cones meet, correspond to pure states with $\abs{h_i} = \frac{1}{2 L}$ for $i = 1,2$, and the remaining points are mixed states for which at least one $h_i$ has absolute value less than $(2L)^{-1}$. The four extreme points marked with black dots in the figure correspond to the massive vacuum $\ket{0_\text{m}}$ and its three zero-momentum excitations. The origin, marked with a white dot, corresponds to the zero-temperature state.

For a quasi-free state, all expectation values can be written in terms of the two-point function $\omega\left( \psi(f) \left[ \psi(g) \right]^\dagger \right)$, which we now compute. For antiperiodic boundary conditions, from Eqs.~\eqref{eq:fourier_ns} and~\eqref{eq:psi_in_annihilation_creation} we find that
\begin{equation}
\label{eq:G_antiperiodic_1}
\omega_\text{NS}\left( \psi(f) \left[ \psi(g) \right]^\dagger \right) = \sum_{n=0}^\infty \left[ \tilde{f}_1\left( - n - \frac{1}{2} \right) \widetilde{g^*_1}\left( n + \frac{1}{2} \right) + \tilde{f}_2\left( n + \frac{1}{2} \right) \widetilde{g^*_2}\left( - n - \frac{1}{2} \right) \right] \eqend{,}
\end{equation}
where we used that $[ \tilde{f}_a(n) ]^* = \widetilde{f^*_a}(-n)$. Since the Fourier transform~\eqref{eq:fourier_def} maps $L^2([0,L])$ to $\ell^2(\mathbb{N})$, the Fourier coefficients $\tilde{f}_a(n)$ are guaranteed to decay faster than $\abs{n}^{-\frac{1}{2}}$, such that the sum~\eqref{eq:G_antiperiodic} is convergent. To determine the integral kernel which we need to compute the modular Hamiltonian using Eq.~\eqref{eq:formula_hg} or the modular flow using Eq.~\eqref{eq:formula_kg}, it is nevertheless useful to introduce a convergence factor $\mathe^{- \frac{\pi}{L} \epsilon (2n+1)}$ and take the limit $\epsilon \to 0^+$ at the end. Using the Fourier transform~\eqref{eq:fourier_def} and noting that we can interchange summation and integral for finite $\epsilon$, we obtain
\begin{splitequation}
\label{eq:G_antiperiodic}
\omega_\text{NS}\left( \psi(f) \left[ \psi(g) \right]^\dagger \right) &= \frac{1}{L} \lim_{\epsilon \to 0^+} \iint_0^L f_1(x) g^*_1(y) \, \sum_{n=0}^\infty \mathe^{- \frac{\pi}{L} \epsilon (2n+1)} \mathe^{- \mathi \frac{\pi (2 n + 1)}{L} (x-y)} \total x \total y \\
&\quad+ \frac{1}{L} \lim_{\epsilon \to 0^+} \iint_0^L f_2(x) g^*_2(y) \, \sum_{n=0}^\infty \mathe^{- \frac{\pi}{L} \epsilon (2n+1)} \mathe^{\mathi \frac{\pi (2 n + 1)}{L} (x-y)} \total x \total y \\
&= \frac{1}{2 \mathi L} \lim_{\epsilon \to 0^+} \iint_0^L \left[ \frac{f_1(x) g^*_1(y)}{\sin\left[ \frac{\pi}{L} (x-y - \mathi \epsilon) \right]} - \frac{f_2(x) g^*_2(y)}{\sin\left[ \frac{\pi}{L} (x-y + \mathi \epsilon) \right]} \right] \total x \total y \eqend{,}
\end{splitequation}
which is the first result of Theorem~\ref{thm:twopf}.

For periodic boundary conditions, we analogously employ Eqs.~\eqref{eq:fourier_r}, \eqref{eq:psi_in_annihilation_creation} and~\eqref{eq:fourier_def} to obtain
\begin{splitequation}
\label{eq:G_periodic}
&\omega_{\text{R},h}\left( \psi(f) \left[ \psi(g) \right]^\dagger \right) = \sum_{a,b=1}^2 L g_{ab} \, \tilde{f}_a(0) \widetilde{g^*_b}(0) + \sum_{n=1}^\infty \left[ \tilde{f}_1(-n) \widetilde{g^*_1}(n) + \tilde{f}_2(n) \widetilde{g^*_2}(-n) \right] \\
&\qquad= \sum_{a,b=1}^2 g_{ab} \iint_0^L f_a(x) g^*_b(y) \total x \total y \\
&\qquad\quad+ \frac{1}{L} \lim_{\epsilon \to 0^+} \iint_0^L \sum_{n=1}^\infty \mathe^{- \frac{2 \pi}{L} n \epsilon} \left[ f_1(x) g^*_1(y) \, \mathe^{- \mathi \frac{2 \pi n}{L} (x-y)} + f_2(x) g^*_2(y) \, \mathe^{\mathi \frac{2 \pi n}{L} (x-y)} \right] \total x \total y \\
&\qquad= \sum_{a,b=1}^2 h_{ab} \iint_0^L f_a(x) g^*_b(y) \total x \total y + \frac{1}{2 \mathi L} \lim_{\epsilon \to 0^+} \iint_0^L f_1(x) g^*_1(y) \cot\left[ \frac{\pi}{L} (x-y - \mathi \epsilon) \right] \total x \total y \\
&\qquad\quad- \frac{1}{2 \mathi L} \lim_{\epsilon \to 0^+} \iint_0^L f_2(x) g^*_2(y) \cot\left[ \frac{\pi}{L} (x-y + \mathi \epsilon) \right] \total x \total y \eqend{,} \raisetag{2em}
\end{splitequation}
where we used a convergence factor $\mathe^{- \frac{2 \pi}{L} n \epsilon}$. Together with Eq.~\eqref{eq:h_decomp} and the characterization of $h$ in the following paragraph, this is the second result of Theorem~\ref{thm:twopf}.

\section{Resolvent}
\label{sec:resolvent}

In this section, we compute the resolvent of the integral kernel $G$ of the two-point function, seen as a convolution operator on $\mathcal{H} = L^2([-\ell,\ell]) \oplus L^2([-\ell,\ell])$ with $2\ell < L$. Since $G$ is bounded with absolutely continuous spectrum $\sigma(G) \subseteq [0,1]$~\cite[Lemma~3.2]{araki1970}, the resolvent $R_G(\mu) = (G - \mu \1)^{-1}$ is an analytic function of the parameter $\mu \in \mathbb{C} \setminus [0,1]$. To determine the integral kernel of $R_G(\mu)$, we first write the resolvent equation in the form
\begin{equation}
\label{eq:reseq}
R_G(\mu) f = \frac{1}{\mu} \Big[ G R_G(\mu) f - f \Big]
\end{equation}
for $f \in \mathcal{H}$, and thus have to determine the unique function $g = R_G(\mu) f$ that solves this equation. Our strategy to do so is to turn Eq.~\eqref{eq:reseq} into a generalized Riemann--Hilbert problem, where one has to determine an analytic function on the cylinder $\mathcal{C}_L = \mathbb{S}^1 \times \mathbb{R}$ satisfying a certain jump condition at the interval $[-\ell,\ell]$. The Riemann--Hilbert problem can then be solved constructively, and we note that this method was first devised by Carleman~\cite{carleman1922} (see also the review~\cite{estradakanwal1987}).

We start from the observation that for any $f \in \mathcal{H}$, $G f$ is the boundary value of the function $H f$, defined on $\mathcal{C}_L \setminus [-\ell,\ell]$ by
\begin{equation}
\label{eq:H_def_NS}
\left( H^\text{NS} f \right)_a(z) \equiv \frac{1}{2 \mathi L} \int_{-\ell}^\ell \frac{\delta_{a1} f_1(y) - \delta_{a2} f_2(y)}{\sin\left[ \frac{\pi}{L} (z-y) \right]} \total y
\end{equation}
for antiperiodic boundary conditions, or
\begin{equation}
\label{eq:H_def_R}
\left( H^\text{R} f \right)_a(z) \equiv \int_{-\ell}^\ell \left[ \sum_{b=1}^2 h_{ab} f_b(y) + \frac{1}{2 \mathi L} \frac{\delta_{a1} f_1(y) - \delta_{a2} f_2(y)}{\tan\left[ \frac{\pi}{L} (z-y) \right]} \right] \total y
\end{equation}
for periodic boundary conditions. Indeed, it is clear from Eqs.~\eqref{eq:G_antiperiodic} and \eqref{eq:G_periodic} that for both boundary conditions we have
\begin{equation}
\label{eq:GH_limit}
(Gf)_1(x) = (Hf)_1^-(x) \equiv \lim_{y \to 0^-} (Hf)_1(x+\mathi y) \eqend{,} \quad (Gf)_2 = (Hf)_2^+ \equiv \lim_{y \to 0^+} (Hf)_2(x+\mathi y) \eqend{,}
\end{equation}
and also in the following we will denote the limit as the interval $[-\ell,\ell]$ is approached from above (resp. below) by a superscript $+$ (resp. $-$), and we will not write the superscript R or NS if an equation holds for both boundary conditions.

This function has the following properties:
\begin{enumerate}
\item[(i)$^\text{R}$] $H^\text{R} f$ is analytic on $\mathcal{C}_L \setminus [-\ell,\ell]$.
\item[(i)$^\text{NS}$] $H^\text{NS} f$ is analytic on the double cover of $\mathcal{C}_L \setminus [-\ell,\ell]$, with $\left( H^\text{NS} f \right)(z+L) = - \left( H^\text{NS} f \right)(z)$.
\item[(ii)] $\displaystyle\lim_{z \to \pm \ell} \Big[ (z \mp \ell) (Hf)(z) \Big] = 0$, where the limit is taken in $\mathcal{C}_L \setminus M_{\ell,\alpha}$ for arbitrary but fixed angle $\alpha \in \left( 0, \frac{\pi}{2} \right)$. Here, $M_{\ell,\alpha} \equiv \{ z \colon \abs{\Re z} + \abs{\Im z} \cot \alpha \leq \ell \}$ is a rhombus with angle $2 \alpha$ at the points $z = \pm \ell$.
\item[(iii)] $(Hf)^- - (Hf)^+ = (f_1, -f_2)$.
\item[(iv)$^\text{NS}$] $\displaystyle\lim_{y \to \infty} \left( H^\text{NS} f \right)(x+\mathi y) = \displaystyle\lim_{y \to \infty} \left( H^\text{NS} f \right)(x-\mathi y) = 0$.
\item[(iv)$^\text{R}$] $\displaystyle\lim_{y \to \infty} \Big[ \hat{h}^+ \left( H^\text{R} f \right)(x+\mathi y) - \hat{h}^- \left( H^\text{R} f \right)(x-\mathi y) \Big] = 0$, where $\hat{h}^\pm \equiv \begin{pmatrix} h_{11} \pm \frac{1}{2L} & - h_{12} \\ h_{21} & - h_{22} \pm \frac{1}{2L} \end{pmatrix}$.
\end{enumerate}
Property (i) is obvious. We prove property (ii) explicitly for $\left( H^\text{NS} f \right)_1$ and $z \to \ell$, since the other cases follow analogously. That is, we want to show that
\begin{equation}
\label{eq:h_property_2_int}
\lim_{z \to \ell} \int_{-\ell}^\ell \frac{(z-\ell) f_1(y)}{\sin\left[ \frac{\pi}{L} (z-y) \right]} \total y = 0 \eqend{,}
\end{equation}
where $z \in \mathcal{C}_L \setminus M_{\ell,\alpha}$ with $\alpha \in \left( 0, \frac{\pi}{2} \right)$. Consider the triangle with corners $z$, $y$ and $\ell$. If $\Re z \geq \ell$, the angle $\theta$ opposite the $z$-$y$ edge is larger or equal to $\frac{\pi}{2}$, and the law of cosines yields
\begin{equation}
\abs{z-y}^2 = \abs{z-\ell}^2 + \abs{y-\ell}^2 - 2 \abs{z-\ell} \abs{y-\ell} \cos \theta \geq \abs{z-\ell}^2 \eqend{.}
\end{equation}
On the other hand, if $\Re z \leq \ell$ we have the bound $\alpha \leq \theta \leq \frac{\pi}{2}$, and the law of sines yields
\begin{equation}
\frac{\abs{z-y}}{\abs{z-\ell}} = \frac{\sin \theta}{\sin \phi} \geq \sin \theta \geq \sin \alpha \eqend{,}
\end{equation}
where $\phi$ is the angle opposite the $z$-$\ell$ edge. So in all cases we have $\abs{z-y} \geq \abs{z-\ell} \sin \alpha$. It follows that the integrand of~\eqref{eq:h_property_2_int} is bounded in absolute value by
\begin{equation}
\label{eq:h_property_2_bound}
\abs{ \frac{(z-\ell) f_1(y)}{\sin\left[ \frac{\pi}{L} (z-y) \right]} } \leq \frac{L}{\pi \sin \alpha} \abs{ f_1(y) } \abs{ \frac{\frac{\pi}{L} (z-y)}{\sin\left[ \frac{\pi}{L} (z-y) \right]} } \eqend{,}
\end{equation}
and we only need to estimate the last term. For this, we use that $\frac{z}{\sin(z)}$ is an analytic function for all $z \neq k \pi$, $k \in \mathbb{Z} \setminus \{ 0 \}$, such that by the maximum principle its absolute value takes its maximum on the boundary of any simply connected bounded region that does not include these points. For $y \in [-\ell,\ell]$ and $z$ close to $\ell$, we can take the region $\abs{ \Im\left[ \frac{\pi}{L} (z-y) \right] } \leq \pi$, $\abs{ \Re\left[ \frac{\pi}{L} (z-y) \right] } \leq a \pi$ with $a = \frac{2 \ell}{L} < 1$, and thus have to bound $\abs{\frac{z}{\sin(z)}}$ on the boundary of this region. For $\Im z = \pm \pi$ and $x = \Re z \in [ - a \pi, a \pi ]$, we obtain
\begin{equation}
\abs{ \frac{z}{\sin(z)} }^2 = \frac{2 (x^2+\pi^2)}{\cosh(2\pi) - \cos(2x)} \leq \frac{2 \pi^2 (a^2+1)}{\cosh(2\pi) - 1} \leq 1 \eqend{,}
\end{equation}
while for $\Re z = \pm a \pi$ and $y = \Im z \in [-\pi,\pi]$ we have
\begin{equation}
\abs{ \frac{z}{\sin(z)} }^2 = \frac{2 (a^2 \pi^2+y^2)}{\cosh(2 y) - \cos(2 a \pi)} \leq \frac{2 \pi^2 (a^2+1)}{1 - \cos(2 a \pi)} \eqend{.}
\end{equation}
It follows that $\abs{ \frac{\frac{\pi}{L} (z-y)}{\sin\left[ \frac{\pi}{L} (z-y) \right]} }$ is bounded by some constant $c = c(a)$ for $z$ close to $\ell$ and $y \in [-\ell,\ell]$. Since $f \in L^2([-\ell,\ell])$ implies $f \in L^1([-\ell,\ell])$, the integrand is thus bounded by an integrable function for all $z$, and we can interchange limit and integral by the dominated convergence theorem, which shows that~\eqref{eq:h_property_2_int} and thus property (ii) holds. The jump condition, property (iii), follows from the Sokhotski--Plemelj formula
\begin{equation}
\label{eq:sokhotski_plemelj}
\lim_{\epsilon \to 0^+} \int \frac{f(y)}{x - y \pm \mathi \epsilon} \total y = \pf \int \frac{f(y)}{x-y} \mp \mathi \pi f(x) \eqend{,}
\end{equation}
where $\pf$ denotes the Cauchy principal value integral, together with the fact that $\frac{z}{\sin(z)} \to 1$ as $z \to 0$. The last property (iv) follows straightforwardly from the definitions, using that for $\Im z \neq 0$ the integrands are bounded in absolute value such that we may again interchange limit and integral by employing the dominated convergence theorem.

\begin{figure}[ht]
\centering
\includegraphics[scale=0.6]{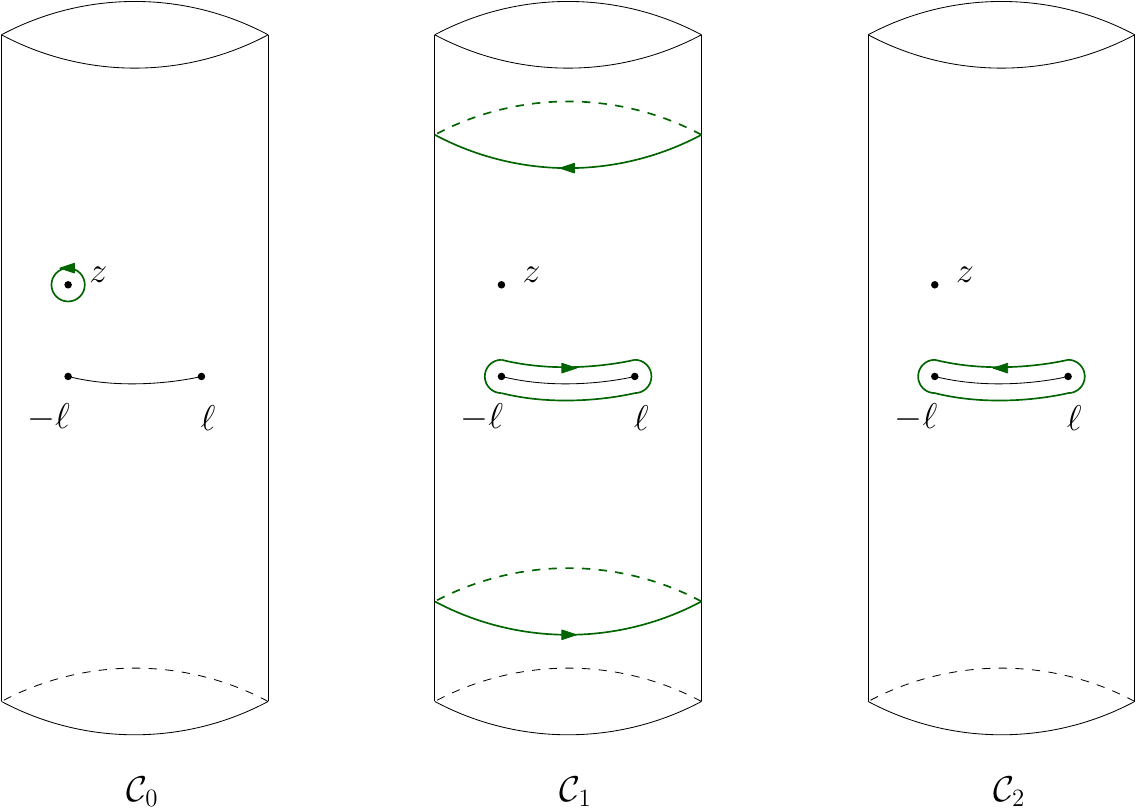}
\caption{Contours of integration used to prove uniqueness of a function that satisfies properties (i)--(iv) on the cylinder $\mathcal{C}_L$. For antiperiodic boundary conditions, one has to consider the double cover of $\mathcal{C}_L$ (not shown).}
\label{fig:contour}
\end{figure}

In fact, the converse is also true: if a function $F$ satisfies the properties listed above, then $F = H f$. Let us show this again explicitly for $H^\text{NS}$, and leave the computation for $H^\text{R}$ to the reader. Consider a point $z \in \mathcal{C}_L \setminus [-\ell,\ell]$. By the Cauchy integral formula for analytic functions, using property (i), we have
\begin{equation}
\label{eq:F_int_C0}
F(z) = - \frac{1}{2 \mathi L} \oint_{\mathcal{C}_0} \frac{F(w)}{\sin\left[ \frac{\pi}{L} (z-w) \right]} \total w \eqend{,}
\end{equation}
where the integration contour $\mathcal{C}_0$ is shown in Fig.~\ref{fig:contour}. Since $F$ is analytic, Cauchy's theorem then shows that we can deform the integration contour into the contour $\mathcal{C}_1$ also shown in Fig.~\ref{fig:contour}. Using property (iv), the contribution of the upper and lower circles vanishes if we send them to $\pm \infty$, and a change of the integration direction gives
\begin{equation}
\label{eq:F_int_C2}
F(z) = \frac{1}{2 \mathi L} \oint_{\mathcal{C}_2} \frac{F(w)}{\sin\left[ \frac{\pi}{L} (z-w) \right]} \total w \eqend{,}
\end{equation}
where the contour $\mathcal{C}_2$ is also shown in Fig.~\ref{fig:contour}. We then split $\mathcal{C}_2$ in four parts: the straight lines with $\abs{\Re w} \leq \ell$ above and below the interval $[-\ell,\ell]$, and two semicircles around the endpoints at $w = \pm \ell$. Consider the semicircle around $z = \ell$, which we parametrize as $w = \ell - \mathi r \mathe^{\mathi t}$ for $t \in [0,\pi]$ and some $r > 0$. Its contribution to the integral~\eqref{eq:F_int_C2} reads
\begin{equation}
\frac{1}{2 L} \int_0^\pi \frac{- \mathi r \mathe^{\mathi t} F(\ell - \mathi r \mathe^{\mathi t})}{\sin\left[ \frac{\pi}{L} (z-\ell + \mathi r \mathe^{\mathi t}) \right]} \total t \eqend{.}
\end{equation}
For small enough $r$, namely such that $r < \abs{z-\ell + k L}$ for all $k \in \mathbb{Z}$, the integrand is bounded in absolute value and again using dominated convergence we can take the limit $r \to 0$ inside the integral. Since $\lim_{r \to 0} [ - \mathi r \mathe^{\mathi t} F(\ell - \mathi r \mathe^{\mathi t}) ] = 0$ by property (ii) (where we can take $\alpha = \frac{\pi}{2}$), the contribution of the semicircle around $z = \ell$ vanishes in the limit where its radius goes to zero. The analogue computation establishes this also for the semicircle around $z = - \ell$. Eq.~\eqref{eq:F_int_C2} thus reduces to the contribution of the straight lines in the limit where they approach the interval:
\begin{equation}
F(z) = \frac{1}{2 \mathi L} \lim_{r \to 0} \int_{-\ell}^\ell \left[ \frac{F(- \mathi r + t)}{\sin\left[ \frac{\pi}{L} (z+ \mathi r - t) \right]} - \frac{F(\mathi r + t)}{\sin\left[ \frac{\pi}{L} (z-\mathi r - t) \right]} \right] \total t \eqend{.}
\end{equation}
For small enough $r$, namely such that $r < \abs{z-t + k L}$ for all $k \in \mathbb{Z}$ and $t \in [-\ell,\ell]$, the integrand is bounded in absolute value and using dominated convergence we can take the limit $r \to 0$ inside the integral. Property (iii) then results in
\begin{equation}
F(z) = \frac{1}{2 \mathi L} \int_{-\ell}^\ell \frac{1}{\sin\left[ \frac{\pi}{L} (z-t) \right]} \begin{pmatrix} f_1(t) \\ - f_2(t) \end{pmatrix} \total t \eqend{,}
\end{equation}
and we see that $F = H^\text{NS} f$ as defined by Eq.~\eqref{eq:H_def_NS}. The analogous computation, starting from
\begin{equation}
F(z) = - \frac{1}{2} \oint_{\mathcal{C}_0} \left[ \frac{1}{\mathi L} \cot\left[ \frac{\pi}{L} (z-w) \right] + \hat{h}_+ + \hat{h}_- \right] F(w) \total w \eqend{,}
\end{equation}
instead of Eq.~\eqref{eq:F_int_C0} and using the relation $L ( \hat{h}_+ - \hat{h}_- ) = \1$, establishes this uniqueness also for $H^\text{R} f$.

If we now define $S \equiv H R_G(\mu) f$ and use the limit~\eqref{eq:GH_limit} in the resolvent equation~\eqref{eq:reseq}, we obtain
\begin{equation}
\label{eq:RS_result}
\big[ R_G(\mu) f \big]_1 = \frac{1}{\mu} \left( S_1^- - f_1 \right) \eqend{,} \quad \big[ R_G(\mu) f \big]_2 = \frac{1}{\mu} \left( S_2^+ - f_2 \right) \eqend{,}
\end{equation}
and property (iii) of $H f$ leads to the jump conditions
\begin{equation}
\label{eq:S_jump}
S_1^- - S_1^+ = \big[ R_G(\mu) f \big]_1 = \frac{1}{\mu} \left( S_1^- - f_1 \right) \eqend{,} \quad S_2^- - S_2^+ = - \big[ R_G(\mu) f \big]_2 = - \frac{1}{\mu} \left( S_2^+ - f_2 \right) \eqend{.}
\end{equation}
Since $S$ also fulfills properties (i), (ii) and (iv), one can again show that it is unique, using in addition that $G-\mu$ is invertible for $\mu \in \mathbb{C} \setminus \sigma(G) = \mathbb{C} \setminus [0,1]$. The problem of finding $S$ is the Riemann--Hilbert problem we alluded to above.

The solution of this Riemann--Hilbert problem is more complicated than of a standard one, since $S$ appears on the right-hand side of the jump condition~\eqref{eq:S_jump} such that the explicit value of the jump is not known. To convert the problem into a standard one, we need to find a partiular solution of the jump condition with $f = 0$. This is however easy: the function
\begin{equation}
\label{eq:rhok_def}
\rho_k(z) \equiv \frac{1}{2 \pi \mathi} \int_{-\ell}^\ell \frac{k(z,y)}{z-y} \total y
\end{equation}
with a continuous function $k$ such that $k(x,x) = 1$ satisfies $\rho_k^- - \, \rho_k^+ = 1$ by the Sokhotski--Plemelj formula~\eqref{eq:sokhotski_plemelj}. Consequently, the function $M_k \equiv \left( 1 - \frac{1}{\mu} \right)^{\rho_k}$ satisfies the jump condition
\begin{equation}
\label{eq:jumpM}
\frac{M_k^-}{M_k^+} = \left( 1 - \frac{1}{\mu} \right)^{\rho_k^- - \rho_k^+} = 1 - \frac{1}{\mu} \eqend{,}
\end{equation}
such that for $A_1 \equiv S_1 M_k$ and $A_2 \equiv S_2 M_k^{-1}$, using Eq.~\eqref{eq:S_jump}, we obtain the jump conditions
\begin{equations}[eq:jumpA]
A_1^- - A_1^+ &= - \frac{1}{\mu} M_k^+ f_1 = \frac{1}{1-\mu} M_k^- f_1 \eqend{,} \\
A_2^- - A_2^+ &= \frac{1}{\mu} \frac{f_2}{M_k^-} = - \frac{1}{1-\mu} \frac{f_2}{M_k^+} \eqend{,}
\end{equations}
where $A$ does not appear on the right-hand side anymore.

To set up the full Riemann--Hilbert problem for $A$, we also need to verify the other properties. To obtain an analytic function for $z \in \mathcal{C}_L \setminus [-\ell,\ell]$, property (i), we need to impose that $k(z,y)$ is analytic for these $z$ and $y \in [-\ell,\ell]$, and that $\frac{k(z,y)}{z-y}$ is a periodic function in $z$ with period $L$. For a vanishing limit as $z \to \pm \ell$, property (ii), we need to impose that
\begin{equation}
\label{eq:M_limit}
\abs{z \mp \ell}^p \abs{ M_k(z) } \leq c \eqend{,} \quad \frac{\abs{z \mp \ell}^p}{\abs{ M_k(z) }} \leq c
\end{equation}
for some $0 < p < 1$, $c > 0$ and $z$ close to $\pm \ell$. Namely, instead of the bound~\eqref{eq:h_property_2_bound} we then obtain
\begin{equation}
\abs{ \frac{(z-\ell) M_k(z) f_1(y)}{\sin\left[ \frac{\pi}{L} (z-y) \right]} } \leq \frac{c L}{\pi \sin \alpha} \frac{\abs{ f_1(y) }}{\abs{z-\ell}^p} \eqend{,}
\end{equation}
which remains integrable, and the limit as $z \to \ell$ of $(z-\ell) M_k(z)$ still vanishes for $p < 1$. For antiperiodic boundary conditions, property (iv)$^\text{NS}$ is maintained as written if we impose that $\lim_{y \to \pm \infty} \rho_k(x+\mathi y) = w_\pm < \infty$, but for periodic boundary conditions we change the explicit form of property (iv)$^\text{R}$. Namely, a short computation reveals that
\begin{equation}
\label{eq:infA_1}
\lim_{y \to \infty} \left[ \dhat{h}^+ A(x+\mathi y) - \dhat{h}^- A(x-\mathi y) \right] = 0 \eqend{,}
\end{equation}
where the matrices $\dhat{h}^\pm$ are given by
\begin{equation}
\dhat{h}^\pm = \begin{pmatrix} \hat{h}^\pm_{11} \left( 1 - \frac{1}{\mu} \right)^{- w_\pm} & \hat{h}^\pm_{12} \left( 1 - \frac{1}{\mu} \right)^{w_\pm} \\ \hat{h}^\pm_{21} \left( 1 - \frac{1}{\mu} \right)^{- w_\pm} & \hat{h}^\pm_{22} \left( 1 - \frac{1}{\mu} \right)^{w_\pm} \end{pmatrix} \eqend{.}
\end{equation}
To obtain property (iv)$^\text{R}$ in the original form (with a matrix $\hat{g}^\pm = \begin{pmatrix} g_{11} \pm \frac{1}{2L} & - g_{12} \\ g_{21} & - g_{22} \pm \frac{1}{2L} \end{pmatrix}$ instead of $\hat{h}^\pm$), we need to multiply Eq.~\eqref{eq:infA_1} with another (constant, invertible) matrix. While this can be done for generic $w_\pm$, the resulting expressions are unwieldy except if $w_- = \pm w_+$. In the case $w_- = w_+ \equiv w$, we multiply Eq.~\eqref{eq:infA_1} with the matrix $B$ and obtain property (iv)$^\text{R}$ in the original form with the matrix $g$, both of which are given by
\begin{equation}
B = \begin{pmatrix} \left( 1 - \frac{1}{\mu} \right)^w & 0 \\ 0 & \left( 1 - \frac{1}{\mu} \right)^{- w} \end{pmatrix} \eqend{,} \quad g = \begin{pmatrix} h_{11} & h_{12} \left( 1 - \frac{1}{\mu} \right)^{2 w} \\ h_{21} \left( 1 - \frac{1}{\mu} \right)^{- 2 w} & h_{22} \end{pmatrix} \eqend{.}
\end{equation}
The case $w \equiv w_- = - w_+$ is more complicated, namely we need
\begin{equation}
B = \frac{- 2 L \left[ \left( 1 - \frac{1}{\mu} \right)^w - \left( 1 - \frac{1}{\mu} \right)^{- w} \right] \left( h - \tr h \, \1 \right) + \left[ \left( 1 - \frac{1}{\mu} \right)^w + \left( 1 - \frac{1}{\mu} \right)^{- w} \right] \1}{1 - 4 L^2 h_1 h_2 + \frac{1}{2} ( 1 + 2 L h_1 ) ( 1 + 2 L h_2 ) \left( 1 - \frac{1}{\mu} \right)^{2 w} + \frac{1}{2} ( 1 - 2 L h_1 ) ( 1 - 2 L h_2 ) \left( 1 - \frac{1}{\mu} \right)^{- 2 w}}
\end{equation}
and obtain
\begin{equation}
\label{eq:riemannhilbert_g_def}
g = \frac{2 h - \tr h \, \1 + \frac{( 1 + 2 L h_1 ) ( 1 + 2 L h_2 )}{4 L} \left( 1 - \frac{1}{\mu} \right)^{2 w} \1 - \frac{( 1 - 2 L h_1 ) ( 1 - 2 L h_2 )}{4 L} \left( 1 - \frac{1}{\mu} \right)^{-2 w} \1}{1 - 4 L^2 h_1 h_2 + \frac{1}{2} ( 1 + 2 L h_1 ) ( 1 + 2 L h_2 ) \left( 1 - \frac{1}{\mu} \right)^{2 w} + \frac{1}{2} ( 1 - 2 L h_1 ) ( 1 - 2 L h_2 ) \left( 1 - \frac{1}{\mu} \right)^{- 2 w}} \eqend{,}
\end{equation}
which we expressed (partially) in terms of the eigenvalues $h_1$ and $h_2$ of $h$~\eqref{eq:h_decomp}. The condition on the eigenvalues $\abs{h_i} \leq \frac{1}{2L}$, $i \in \{1,2\}$, guarantees that the denominator of both matrices never vanishes.

We have proved above that the solution of the Riemann--Hilbert problem is unique for a function satisfying properties (i)--(iv), namely it is given by $A = H f$ with $f$ the function appearing on the right-hand side of the jump condition~\eqref{eq:jumpA}. For antiperiodic boundary conditions, $H^\text{NS} f$ is given by Eq.~\eqref{eq:H_def_NS} such that
\begin{equation}
A(z) = \frac{1}{2 \mathi L} \frac{1}{1-\mu} \int_{-\ell}^\ell \frac{1}{\sin\left[ \frac{\pi}{L} (z-y) \right]} \begin{pmatrix} M_k^-(y) f_1(y) \\ \frac{f_2(y)}{M_k^+(y)} \end{pmatrix} \total y \eqend{,}
\end{equation}
and expressing this using $S(z)$ and taking the limit from above and below the interval, we obtain the resolvent~\eqref{eq:RS_result}
\begin{equations}[eq:RNS_f]
\left[ R^\text{NS}_G(\mu) f \right]_1(x) &= \frac{1}{2 \mathi L} \frac{1}{\mu (1-\mu)} \lim_{\epsilon \to 0^+} \int_{-\ell}^\ell \frac{f_1(y)}{\sin\left[ \frac{\pi}{L} (x-y - \mathi \epsilon) \right]} \frac{M_k(y - \mathi \epsilon)}{M_k(x - \mathi \epsilon)} \total y - \frac{1}{\mu} f_1(x) \eqend{,} \\
\left[ R^\text{NS}_G(\mu) f \right]_2(x) &= \frac{1}{2 \mathi L} \frac{1}{\mu (1-\mu)} \lim_{\epsilon \to 0^+} \int_{-\ell}^\ell \frac{f_2(y)}{\sin\left[ \frac{\pi}{L} (x-y + \mathi \epsilon) \right]} \frac{M_k(x + \mathi \epsilon)}{M_k(y + \mathi \epsilon)} \total y - \frac{1}{\mu} f_2(x) \eqend{.}
\end{equations}
It only remains to determine the function $k$ in the definition~\eqref{eq:rhok_def} of $\rho_k$ to obtain the explicit expression for $M_k = \left( 1 - \frac{1}{\mu} \right)^{\rho_k}$. Since $\frac{k(z,y)}{z-y}$ must be analytic for $z \in \mathcal{C}_L \setminus [-\ell,\ell]$ and periodic in $z$ to fulfill property (i), and satisfy $k(z,z) = 1$, we have the unique choice $k(z,y) = \frac{\pi}{L} (z-y) \cot\left[ \frac{\pi}{L} (z-y) \right]$. It then follows that
\begin{equation}
\label{eq:rho_result}
\rho_k(z) = \frac{1}{2 \mathi L} \int_{-\ell}^\ell \cot\left[ \frac{\pi}{L} (z-y) \right] \total y = \frac{\mathi}{2 \pi} \ln\left[ \frac{\sin\left[ \frac{\pi}{L} (z-\ell) \right]}{\sin\left[ \frac{\pi}{L} (z+\ell) \right]} \right] \eqend{,}
\end{equation}
and we compute
\begin{equation}
\label{eq:rho_boundary}
\rho_k^\pm(z) = \lim_{y \to 0^+} \rho_k(x \pm \mathi y) = - \frac{\mathi}{2 \pi} \Omega_1(x) \mp \frac{1}{2} \eqend{.}
\end{equation}
with the function~\eqref{eq:flow_omega_def}
\begin{equation}
\label{eq:omega_1_def}
\Omega_1(x) \equiv \ln\left( \frac{\sin\left[ \frac{\pi}{L} (\ell+x) \right]}{\sin\left[ \frac{\pi}{L} (\ell-x) \right]} \right) \eqend{.}
\end{equation}

We also verify that $\lim_{y \to \pm \infty} \rho_k(x + \mathi y) = \mp \frac{\ell}{L}$, and that 
\begin{splitequation}
\abs{ M_k(z) } &= \abs{ \left( 1 - \frac{1}{\mu} \right)^{\frac{\mathi}{2 \pi} \ln\left[ \frac{\sin\left[ \frac{\pi}{L} (z-\ell) \right]}{\sin\left[ \frac{\pi}{L} (z+\ell) \right]} \right]} } = \abs{ \left[ \frac{\sin\left[ \frac{\pi}{L} (z-\ell) \right]}{\sin\left[ \frac{\pi}{L} (z+\ell) \right]} \right]^{\frac{\mathi}{2 \pi} \ln\left( 1 - \frac{1}{\mu} \right)} } \\
&\leq c \abs{ z - \ell }^{- \frac{1}{2 \pi} \arg\left( 1 - \frac{1}{\mu} \right)} \exp\left[ - \frac{1}{2 \pi} \ln\abs{ 1 - \frac{1}{\mu} } \arg(z - \ell) \right] \\
&\leq c \abs{ z - \ell }^{- \frac{1}{2}} \exp\left[ \frac{1}{2 \pi} \abs{ \ln\abs{ 1 - \frac{1}{\mu} } } (\pi-\alpha) \right]
\end{splitequation}
for $z \in \mathcal{C}_L \setminus M_{\ell,\alpha}$ and close to $\ell$, and some constant $c$ depending on $\ell$, $L$ and $\mu$. Analogous expressions are obtained for $z$ close to $-\ell$ and for $1/\abs{M_k(z)}$, such that the conditions~\eqref{eq:M_limit} are seen to hold with $p = \frac{1}{2}$. Inserting the result~\eqref{eq:rho_boundary} into Eq.~\eqref{eq:RNS_f}, we then can finally read off the resolvent operator for antiperiodic boundary conditions:
\begin{equation}
\left[ R^\text{NS}_G(\mu) f \right]_a(x) = \sum_{b=1}^2 \int_{-\ell}^\ell R^\text{NS}_{ab}(\mu; x,y) f_b(y) \total y \eqend{,}
\end{equation}
with the distributional integral kernel
\begin{splitequation}
\label{eq:resolvent_ns_kernel}
R^\text{NS}_{ab}(\mu; x,y) &= \frac{1}{\mu (1-\mu)} \frac{1}{2 \mathi L} \lim_{\epsilon \to 0^+} \begin{pmatrix} \frac{1}{\sin\left[ \frac{\pi}{L} (x-y - \mathi \epsilon) \right]} & 0 \\ 0 & - \frac{1}{\sin\left[ \frac{\pi}{L} (x-y + \mathi \epsilon) \right]} \end{pmatrix}_{ab} \left( 1 - \frac{1}{\mu} \right)^{\frac{\mathi}{2 \pi} \left[ \Omega_a(x) - \Omega_b(y) \right]} \\
&\quad- \frac{1}{\mu} \delta_{ab} \delta(x-y) \\
&= \frac{1}{\mu (1-\mu)} \frac{1}{2 \mathi L} \begin{pmatrix} 1 & 0 \\ 0 & - 1 \end{pmatrix}_{ab} \left( 1 - \frac{1}{\mu} \right)^{\frac{\mathi}{2 \pi} \left[ \Omega_a(x) - \Omega_b(y) \right]} \pf \frac{1}{\sin\left[ \frac{\pi}{L} (x-y) \right]} \\
&\quad- \frac{1 - 2 \mu}{2 \mu (1-\mu)} \delta_{ab} \delta(x-y) \eqend{,} \raisetag{2.2em}
\end{splitequation}
and where in addition to~\eqref{eq:omega_1_def} we also defined
\begin{equation}
\label{eq:omega_2_def}
\Omega_2(x) \equiv - \Omega_1(x) = \Omega_1(-x) \eqend{.}
\end{equation}

For periodic boundary conditions, since we have the limit $w_\pm = \lim_{y \to \pm \infty} \rho_k(x + \mathi y) = \mp \frac{\ell}{L}$, we are in the situation with $w_- = - w_+$. It follows that the solution of the Riemann--Hilbert problem for $A$ is given by $A = H^\text{R} f$~\eqref{eq:H_def_R} with $f$ the function appearing on the right-hand side of the jump condition~\eqref{eq:jumpA}, and with the matrix $h$ replaced by $g$ defined in Eq.~\eqref{eq:riemannhilbert_g_def} with $w = \frac{\ell}{L}$. This reads explicitly
\begin{equation}
A_a(z) = \frac{1}{1-\mu} \int_{-\ell}^\ell \left[ g_{a1} M_k^-(y) f_1(y) + g_{a2} \frac{f_2(y)}{M_k^+(y)} + \frac{\delta_{a1} M_k^-(y) f_1(y) - \delta_{a2} \frac{f_2(y)}{M_k^+(y)}}{2 \mathi L \tan\left[ \frac{\pi}{L} (z-y) \right]} \right] \total y \eqend{,}
\end{equation}
and expressing $A(z)$ using $S(z)$ and taking the limit from above and below the interval, we obtain the resolvent~\eqref{eq:RS_result}
\begin{equations}[eq:RR_f]
\begin{split}
\left[ R^\text{R}_G(\mu) f \right]_1(x) &= \frac{1}{\mu (1-\mu)} \lim_{\epsilon \to 0^+} \int_{-\ell}^\ell \frac{1}{M_k(z - \mathi \epsilon)} \bigg[ g_{11}(\mu) M_k(y - \mathi \epsilon) f_1(y) + g_{12}(\mu) \frac{f_2(y)}{M_k(y + \mathi \epsilon)} \\
&\hspace{10em}+ \frac{M_k(y - \mathi \epsilon) f_1(y)}{2 \mathi L \tan\left[ \frac{\pi}{L} (z-y - \mathi \epsilon) \right]} \bigg] \total y - \frac{1}{\mu} f_1(x) \eqend{,}
\end{split} \raisetag{5.4em} \\
\begin{split}
\left[ R^\text{R}_G(\mu) f \right]_2(x) &= \frac{1}{\mu (1-\mu)} \lim_{\epsilon \to 0^+} \int_{-\ell}^\ell M_k(z + \mathi \epsilon) \bigg[ g_{21}(\mu) M_k(y - \mathi \epsilon) f_1(y) + g_{22}(\mu) \frac{f_2(y)}{M_k(y + \mathi \epsilon)} \\
&\hspace{10em}- \frac{\frac{f_2(y)}{M_k(y + \mathi \epsilon)}}{2 \mathi L \tan\left[ \frac{\pi}{L} (z-y + \mathi \epsilon) \right]} \bigg] \total y - \frac{1}{\mu} f_2(x) \eqend{,}
\end{split} \raisetag{2.2em}
\end{equations}
where we also explicitly indicated the dependence of the matrix $g$~\eqref{eq:riemannhilbert_g_def} on $\mu$. Inserting the solution~\eqref{eq:rho_result} for $\rho_k$, and employing the result~\eqref{eq:rho_boundary} for its boundary values, we can finally read off the resolvent operator for periodic boundary conditions:
\begin{equation}
\left[ R^\text{R}_G(\mu) f \right]_a(x) = \sum_{b=1}^2 \int_{-\ell}^\ell R^\text{R}_{ab}(\mu; x,y) f_b(y) \total y \eqend{,}
\end{equation}
with the distributional integral kernel
\begin{splitequation}
\label{eq:resolvent_r_kernel}
R^\text{R}_{ab}(\mu; x,y) &= \frac{1}{\mu (1-\mu)} \bigg[ \frac{1}{2 \mathi L} \lim_{\epsilon \to 0^+} \begin{pmatrix} \cot\left[ \frac{\pi}{L} (x-y - \mathi \epsilon) \right] & 0 \\ 0 & - \cot\left[ \frac{\pi}{L} (x-y + \mathi \epsilon) \right] \end{pmatrix}_{ab} \\
&\hspace{6em}+ g_{ab}(\mu) \bigg] \left( 1 - \frac{1}{\mu} \right)^{\frac{\mathi}{2\pi} \left[ \Omega_a(x) - \Omega_b(y) \right]} - \frac{1}{\mu} \delta_{ab} \delta(x-y) \\
&= \frac{1}{\mu (1-\mu)} \bigg[ \frac{1}{2 \mathi L} \begin{pmatrix} 1 & 0 \\ 0 & - 1 \end{pmatrix}_{ab} \pf \cot\left[ \frac{\pi}{L} (x-y) \right] + g_{ab}(\mu) \bigg] \left( 1 - \frac{1}{\mu} \right)^{\frac{\mathi}{2\pi} \left[ \Omega_a(x) - \Omega_b(y) \right]} \\
&\quad- \frac{1-2\mu}{2 \mu (1-\mu)} \delta_{ab} \delta(x-y) \eqend{.} \raisetag{2em}
\end{splitequation}

For large $\abs{\mu}$ it is also possible to compute the resolvent via the Neumann series
\begin{equation}
\label{eq:resolvent_neumann}
R_A(\mu) = (A - \mu \1)^{-1} = - \frac{1}{\mu} \sum_{n=0}^\infty \frac{A^n}{\mu^n} = - \frac{1}{\mu} - \frac{1}{\mu^2} A + \bigo{\mu^{-3}} \eqend{,}
\end{equation}
which is absolutely convergent for $\abs{\mu} > \norm{ A }$. Comparing the results~\eqref{eq:resolvent_ns_kernel} and~\eqref{eq:resolvent_r_kernel} with the integral kernels of the two-point functions~\eqref{eq:G_periodic} and~\eqref{eq:G_antiperiodic} and using that $g = h + \bigo{\mu^{-1}}$, which follows from the explicit expression~\eqref{eq:riemannhilbert_g_def}, we find agreement with the Neumann series up to order $\mu^{-2}$. This provides an additional check on the computation.

Lastly, we have to verify what happens in the limit where $\mu \to 0$ or $\mu \to 1$. Since $0$ and $1$ are not eigenvalues of the two-point function $G$ (seen as convolution operator)~\cite[Corollary~4.10]{araki1970}, the resolvent could exist (as an unbounded operator) for these values of $\mu$, but our results~\eqref{eq:resolvent_ns_kernel} and~\eqref{eq:resolvent_r_kernel} do not show this. To compute the limit, we employ Lemma~\ref{lemma:limit} with $t = \frac{1}{2 \pi} \left[ \Omega_1(x) - \Omega_1(y) \right]$~\eqref{eq:omega_1_def}, which is the identity
\begin{equation}
\lim_{a \to 0} \left[ a^{\frac{\mathi}{2 \pi} \left[ \Omega_1(x) - \Omega_1(y) \right]} \pf \frac{1}{\sinh\left[ \frac{1}{2} \Omega_1(x) - \frac{1}{2} \Omega_1(y) \right]} - \mathi \frac{a-1}{a+1} \delta \left[ \frac{\Omega_1(x) - \Omega_1(y)}{2 \pi} \right] \right] = 0 \eqend{.}
\end{equation}
For the first term, we compute that
\begin{splitequation}
\label{eq:resolvent_sinh_trafo}
&\frac{1}{\sinh\left[ \frac{1}{2} \Omega_1(x) - \frac{1}{2} \Omega_1(y) \right]} \\
&\quad= \frac{2 \sqrt{ \sin\left[ \frac{\pi}{L} (\ell+x) \right] \sin\left[ \frac{\pi}{L} (\ell-x) \right] \sin\left[ \frac{\pi}{L} (\ell+y) \right] \sin\left[ \frac{\pi}{L} (\ell-y) \right] }}{\sin\left( \frac{2 \pi}{L} \ell \right)} \frac{1}{\sin\left[ \frac{\pi}{L} (x-y) \right]} \eqend{.}
\end{splitequation}
On the other hand, the well-known composition formula for the Dirac $\delta$ yields
\begin{equation}
\delta\left[ \frac{\Omega_1(x) - \Omega_1(y)}{2 \pi} \right] = \frac{2 L \sin\left[ \frac{\pi}{L} (\ell+x) \right] \sin\left[ \frac{\pi}{L} (\ell-x) \right]}{\sin\left( \frac{2 \pi}{L} \ell \right)} \delta(x-y)
\end{equation}
for the second term, such that we obtain the distributional limit
\begin{equation}
\label{eq:limit_lemma_omega_1}
\lim_{a \to 0} \left[ a^{\frac{\mathi}{2 \pi} \left[ \Omega_1(x) - \Omega_1(y) \right]} \pf \frac{1}{\sin\left[ \frac{\pi}{L} (x-y) \right]} - \mathi L \frac{a-1}{a+1} \delta(x-y) \right] = 0 \eqend{.}
\end{equation}
The analogous computation results in the further limits
\begin{equations}[eq:limit_lemma_omega_2]
\lim_{a \to \infty} \left[ a^{\frac{\mathi}{2 \pi} \left[ \Omega_1(x) - \Omega_1(y) \right]} \pf \frac{1}{\sin\left[ \frac{\pi}{L} (x-y) \right]} - \mathi L \frac{a-1}{a+1} \delta(x-y) \right] &= 0 \eqend{,} \\
\lim_{a \to 0} \left[ a^{\frac{\mathi}{2 \pi} \left[ \Omega_2(x) - \Omega_2(y) \right]} \pf \frac{1}{\sin\left[ \frac{\pi}{L} (x-y) \right]} + \mathi L \frac{a-1}{a+1} \delta(x-y) \right] &= 0 \eqend{,} \\
\lim_{a \to \infty} \left[ a^{\frac{\mathi}{2 \pi} \left[ \Omega_2(x) - \Omega_2(y) \right]} \pf \frac{1}{\sin\left[ \frac{\pi}{L} (x-y) \right]} + \mathi L \frac{a-1}{a+1} \delta(x-y) \right] &= 0 \eqend{,}
\end{equations}
which only differ in signs.

It is an easy exercise to show that for any bounded operator $A$ we have
\begin{equation}
\norm{ ( A - \mu \1 )^{-1} } = \operatorname{dist}(\mu, \sigma(A)) \eqend{,}
\end{equation}
such that the resolvent cannot diverge too strongly as $\mu \to 0,1$, at most like $\mu^{-1}$ (close to $\mu = 0$) or $(1-\mu)^{-1}$ (close to $\mu = 1$). In fact, we can rewrite our result~\eqref{eq:resolvent_ns_kernel} for the resolvent integral kernel for antiperiodic boundary conditions in the form
\begin{equations}[eq:resolvent_ns_kernel_2]
\begin{split}
R^\text{NS}_{11}(\mu; x,y) &= \frac{1}{\mu (1-\mu)} \frac{1}{2 \mathi L} \mathe^{\frac{1}{2 \pi} \left[ \Omega_1(x) - \Omega_1(y) \right] \left[ \arg\left( \frac{1}{\mu} - 1 \right) - \arg\left( 1 - \frac{1}{\mu} \right) \right]} \\
&\qquad\times \left[ \left( \frac{1}{\mu} - 1 \right)^{\frac{\mathi}{2 \pi} \left[ \Omega_1(x) - \Omega_1(y) \right]} \pf \frac{1}{\sin\left[ \frac{\pi}{L} (x-y) \right]} - \mathi L (1-2\mu) \delta(x-y) \right] \eqend{,} \raisetag{5.2em}
\end{split} \\
\begin{split}
R^\text{NS}_{22}(\mu; x,y) &= - \frac{1}{\mu (1-\mu)} \frac{1}{2 \mathi L} \mathe^{\frac{1}{2 \pi} \left[ \Omega_2(x) - \Omega_2(y) \right] \left[ \arg\left( \frac{1}{\mu} - 1 \right) - \arg\left( 1 - \frac{1}{\mu} \right) \right]} \\
&\qquad\times\left[ \left( \frac{1}{\mu} - 1 \right)^{\frac{\mathi}{2 \pi} \left[ \Omega_2(x) - \Omega_2(y) \right]} \pf \frac{1}{\sin\left[ \frac{\pi}{L} (x-y) \right]} + \mathi L (1-2\mu) \delta(x-y) \right] \eqend{,} \raisetag{5.2em}
\end{split}
\end{equations}
and in the limits $\mu \to 0$ and $\mu \to 1$ the terms in brackets on the right-hand side vanish, taking $a = \frac{1}{\mu} - 1$ (which is real for $\mu \in [0,1]$) in the results~\eqref{eq:limit_lemma_omega_1} and~\eqref{eq:limit_lemma_omega_2}. It follows that in our case, the resolvent is actually less divergent than what would be allowed by the general formula, which will be relevant later on. In the same way, we rewrite our result~\eqref{eq:resolvent_r_kernel} for the resolvent integral kernel for periodic boundary conditions in the form
\begin{equations}[eq:resolvent_r_kernel_2]
\begin{split}
R^\text{R}_{11}(\mu; x,y) &= \frac{1}{\mu (1-\mu)} \frac{\cos\left[ \frac{\pi}{L} (x-y) \right]}{2 \mathi L} \mathe^{\frac{1}{2 \pi} \left[ \Omega_1(x) - \Omega_1(y) \right] \left[ \arg\left( \frac{1}{\mu} - 1 \right) - \arg\left( 1 - \frac{1}{\mu} \right) \right]} \\
&\qquad\times \left[ \left( \frac{1}{\mu} - 1 \right)^{\frac{\mathi}{2\pi} \left[ \Omega_1(x) - \Omega_1(y) \right]} \pf \frac{1}{\sin\left[ \frac{\pi}{L} (x-y) \right]} - \mathi L (1-2\mu) \delta(x-y) \right] \\
&\quad+ \frac{1}{\mu (1-\mu)} \left( 1 - \frac{1}{\mu} \right)^{\frac{\mathi}{2\pi} \left[ \Omega_1(x) - \Omega_1(y) \right]} g_{11}(\mu) \eqend{,} \raisetag{2em}
\end{split} \\
R^\text{R}_{12}(\mu; x,y) &= \frac{1}{\mu (1-\mu)} \left( 1 - \frac{1}{\mu} \right)^{\frac{\mathi}{2\pi} \left[ \Omega_1(x) - \Omega_2(y) \right]} g_{12}(\mu) \eqend{,} \\
R^\text{R}_{21}(\mu; x,y) &= \frac{1}{\mu (1-\mu)} \left( 1 - \frac{1}{\mu} \right)^{\frac{\mathi}{2\pi} \left[ \Omega_2(x) - \Omega_1(y) \right]} g_{21}(\mu) \eqend{,} \\
\begin{split}
R^\text{R}_{22}(\mu; x,y) &= - \frac{1}{\mu (1-\mu)} \frac{\cos\left[ \frac{\pi}{L} (x-y) \right]}{2 \mathi L} \mathe^{\frac{1}{2 \pi} \left[ \Omega_2(x) - \Omega_2(y) \right] \left[ \arg\left( \frac{1}{\mu} - 1 \right) - \arg\left( 1 - \frac{1}{\mu} \right) \right]} \\
&\qquad\times\left[ \left( \frac{1}{\mu} - 1 \right)^{\frac{\mathi}{2\pi} \left[ \Omega_2(x) - \Omega_2(y) \right]} \pf \frac{1}{\sin\left[ \frac{\pi}{L} (x-y) \right]} + \mathi L (1-2\mu) \delta(x-y) \right] \\
&\quad- \frac{1}{\mu (1-\mu)} \left( 1 - \frac{1}{\mu} \right)^{\frac{\mathi}{2\pi} \left[ \Omega_2(x) - \Omega_2(y) \right]} g_{22}(\mu) \eqend{.} \raisetag{2em}
\end{split}
\end{equations}
In the limits $\mu \to 0$ and $\mu \to 1$, the terms in brackets again vanish, such that the resolvent is less divergent than what one would expect. For the terms involving the matrix $g$ defined in Eq.~\eqref{eq:riemannhilbert_g_def}, we compute first that
\begin{equation}
g(\mu) = \frac{1}{2 L} \1 + \bigo{ \mu^\frac{2 \ell}{L} } \eqend{,} \quad g(\mu) = - \frac{1}{2 L} \1 + \bigo{ (1-\mu)^\frac{2 \ell}{L} } \eqend{.}
\end{equation}
The off-diagonal components $R^\text{R}_{12}$ and $R^\text{R}_{21}$ of the resolvent integral kernel~\eqref{eq:resolvent_r_kernel_2} thus only diverge $\sim \mu^{1-\frac{2 \ell}{L}}$ as $\mu \to 0$ or $\sim (1-\mu)^{1-\frac{2 \ell}{L}}$ as $\mu \to 1$, which is again less than what would be expected in general. On the other hand, for the diagonal components $g_{aa}$ becomes constant, and multiplying the results~\eqref{eq:limit_lemma_omega_1} and~\eqref{eq:limit_lemma_omega_2} by $\sin\left[ \frac{\pi}{L} (x-y) \right]$ we obtain that
\begin{equation}
\lim_{\mu \to 0} \left( 1 - \frac{1}{\mu} \right)^{\frac{\mathi}{2\pi} \left[ \Omega_a(x) - \Omega_a(y) \right]} = 0 = \lim_{\mu \to 1} \left( 1 - \frac{1}{\mu} \right)^{\frac{\mathi}{2\pi} \left[ \Omega_a(x) - \Omega_a(y) \right]}
\end{equation}
in the weak topology of distributions (this is essentially the Riemann--Lebesgue lemma). Therefore, also the diagonal components $R^\text{R}_{11}$ and $R^\text{R}_{22}$ of the resolvent integral kernel~\eqref{eq:resolvent_r_kernel_2} are less divergent.

\section{Modular flow and modular Hamiltonian}
\label{sec:flow}

In this section, we determine the modular flow and its generator, the modular Hamiltonian. As stated in the introduction, for free theories the modular flow and modular Hamiltonian are second-quantized operators on Fock space~\cite{eckmannosterwalder1973,figlioliniguido1989} and it is enough to determine the single-particle operators. These act as given in Thms.~\ref{thm:flow} and~\ref{thm:hamiltonian}, with the corresponding integral kernels $K$ given by the formulas~\eqref{eq:formula_hg} (for the modular Hamiltonian) and~\eqref{eq:formula_kg} (for the modular flow). The right-hand sides of these formulas can be defined via spectral theory, and the properties of the two-point function ensure that they are well-defined. To perform concrete computations, one then has to choose a functional calculus. For a bounded operator $A$ and a function $f$ that is holomorphic on an open set containing the spectrum $\sigma(A)$ of $A$, the holomorphic functional calculus defines $f(A)$ as the operator
\begin{equation}
\label{eq:calculus_holomorphic_f}
f(A) = \frac{\mathi}{2 \pi} \oint f(z) R_A(z) \total z \eqend{,}
\end{equation}
where $R_A(z) = ( A - z \1 )^{-1}$ is the resolvent of $A$, and the integration contour lies inside the set where $f$ is holomorphic and encloses the spectrum $\sigma(A)$ (with winding number 1). This is the approach that was followed in recent work by Erdmenger et al.~\cite{erdmengerfriesreyessimon2020}. However, neither the function $f(z) = \left( \frac{z}{1-z} \right)^{\mathi t}$ nor $f(z) = \ln \left( \frac{z}{1-z} \right)$, which are the ones that are needed to compute the modular data, are holomorphic on an open set containing the spectrum $[0,1]$ since they have branch cuts starting at $z = 0$ and $z = 1$ and extending outwards. The way that this issue was resolved in~\cite{erdmengerfriesreyessimon2020} is to replace the function $f$ by a regulated version $f_\epsilon$ whose branch cuts start at $z = - \epsilon$ and $z = 1 + \epsilon$, and at the end of the computation take the limit $\epsilon \to 0^+$. If this limit is taken carefully (as done in~\cite{erdmengerfriesreyessimon2020}), one obtains the correct results, but a rigorous mathematical analysis of the space in which this limit takes place and under which conditions it exists was not given there.

On the other hand, for any self-adjoint operator $A$ the spectral theorem~\cite[Theorem~VIII.6]{reedsimon1} shows that there exists a unique projection-valued (spectral) measure $E_A$ whose support is the spectrum $\sigma(A)$. Using this measure, one defines the operator $f(A)$ by its matrix elements
\begin{equation}
\label{eq:calculus_borel_f}
\bra{\phi} f(A) \ket{\phi'} = \int_{\sigma(A)} f(\mu) \total \bra{\phi} E_A(\mu) \ket{\phi'} \eqend{,}
\end{equation}
where $\ket{\phi}$ and $\ket{\phi'}$ are elements of a dense subspace $\mathcal{D}_{f,A}$ of Hilbert space, namely the subspace for which the right-hand side is finite~\cite[Eq.~VIII.5]{reedsimon1}. One can show that this is an extension of the holomorphic functional calculus to (unbounded) Borel functions, i.e., for holomorphic functions and bounded operators both definitions result in the same operator. Since the functions that we need are continuous in the interval $(0,1)$, but diverge at the endpoints $0$ and $1$, this is the right functional calculus to use. If $0$ and $1$ were eigenfunctions of our operator $A$, $\mathcal{D}_{f,A}$ would not be dense (namely, its complement would contain the linear span of the corresponding eigenfunctions). Fortunately, this is excluded by~\cite[Corollary~4.10]{araki1970}, and so we can use Eq.~\eqref{eq:calculus_borel_f} without problems.

For a bounded self-adjoint operator, formula~\eqref{eq:calculus_borel_f} can be further simplified. Namely, Stone's formula~\eqref{eq:stone_formula} for an absolutely continuous spectrum shows that the spectral measure can be defined by~\cite[Theorem~VII.13]{reedsimon1}
\begin{equation}
\total E_A(\mu) = \frac{1}{2 \pi \mathi} \lim_{\epsilon \to 0^+} \Big[ R_A(\mu + \mathi \epsilon) - R_A(\mu - \mathi \epsilon) \Big] \total \mu \eqend{,}
\end{equation}
where the convergence takes place in the strong operator topology, i.e., acting on a vector $\ket{\phi}$ in Hilbert space. We are thus led to consider the limit
\begin{equation}
\label{eq:calculus_resolvent}
f(A) \ket{\phi} \equiv \frac{1}{2 \pi \mathi} \int_{\sigma(A)} f(\mu) \lim_{\epsilon \to 0^+} \Big[ R_A(\mu + \mathi \epsilon) - R_A(\mu - \mathi \epsilon) \Big] \ket{\phi} \total \mu
\end{equation}
for $\ket{\phi}$ in a suitable dense subspace of the Hilbert space. Taking $f(z) = \left( \frac{z}{1-z} \right)^{\mathi t}$, this is nothing else but the formula~\eqref{eq:flow_singleparticle_function}
\begin{equation}
\label{eq:flow_singleparticle_function_2}
f_a(t,x) = \frac{1}{2 \pi \mathi} \int_0^1 \left( \frac{\mu}{1-\mu} \right)^{\mathi t} \lim_{\epsilon \to 0^+} \sum_{b=1}^2 \int_{-\ell}^\ell \Big[ R(\mu + \mathi \epsilon)_{ab}(x,y) - R(\mu - \mathi \epsilon)_{ab}(x,y) \Big] f_b(y) \total y \total \mu \eqend{,}
\end{equation}
for the modular flow of a vector in the single-particle Hilbert space $\mathcal{H}$, i.e., the definition of the function $f_a(t,x)$ in Eq.~\eqref{eq:flow_singleparticle}. Note that the formula~\eqref{eq:calculus_resolvent} is effectively the same as one would obtain from the holomorphic functional calculus, ignoring the possible contribution from the endpoints $z = 0$ and $1$ and only keeping the ones from above and below the cut at $[0,1]$.

In a first step, we determine the difference of resolvents, respectively their integral kernels. For antiperiodic boundary conditions, the integral kernel of the resolvent is given in Eq.~\eqref{eq:resolvent_ns_kernel_2}, and has the general form
\begin{equation}
\label{eq:resolvent_gen}
R(\mu) = \frac{1}{\mu (1-\mu)} \mathe^{a \left[ \arg\left( \frac{1}{\mu} - 1 \right) - \arg\left( 1 - \frac{1}{\mu} \right) \right]} f(\mu) = \left[ \frac{1}{\mu} + \frac{1}{1-\mu} \right] \mathe^{a \left[ \arg\left( \frac{1}{\mu} - 1 \right) - \arg\left( 1 - \frac{1}{\mu} \right) \right]} f(\mu) \eqend{,}
\end{equation}
where $f(\mu)$ is a function that is continuous across the cut $\mu \in [0,1]$. Taking the difference from above and below the cut, we obtain
\begin{splitequation}
\label{eq:resolvent_diff_gen}
\lim_{\epsilon \to 0^+} \Big[ R(\mu + \mathi \epsilon) - R(\mu - \mathi \epsilon) \Big] &= - 2 \sinh(\pi a) f(\mu) \left[ \pf \frac{1}{\mu} + \pf \frac{1}{1-\mu} \right] \\
&\quad- 2 \pi \mathi \cosh(\pi a) \Big[ \delta(\mu) f(\mu) - \delta(1-\mu) f(\mu) \Big] \eqend{,}
\end{splitequation}
where we employed the Sokhotski--Plemelj formula~\eqref{eq:sokhotski_plemelj}. However, as we have seen the terms in brackets in Eq.~\eqref{eq:resolvent_ns_kernel_2} which constitute $f(\mu)$ actually vanish for $\mu = 0$ and $\mu = 1$. It follows that the terms in the second line of Eq.~\eqref{eq:resolvent_diff_gen} vanish, and for the terms in the first line we can dispense with the finite-part prescription. That is, we simply have
\begin{equation}
\label{eq:resolvent_diff_gen_2}
\lim_{\epsilon \to 0^+} \Big[ R(\mu + \mathi \epsilon) - R(\mu - \mathi \epsilon) \Big] = - 2 \sinh(\pi a) \frac{1}{\mu (1-\mu)} f(\mu) \eqend{,}
\end{equation}
which is exactly the result that one would obtain by only taking into account the jump of the resolvent~\eqref{eq:resolvent_gen} across the interval $(0,1)$ and ignoring any possible contribution from the endpoints $\mu = 0$ and $\mu = 1$. This is the prescription that was adopted in~\cite{erdmengerfriesreyessimon2020}, such that our derivation can also serve as a rigorous justification for the computations there. From the above results, one sees that a necessary and sufficient condition for this prescription to work is that
\begin{equation}
\lim_{\mu \to 0} \Big[ \mu R(\mu) \Big] = \lim_{\mu \to 1} \Big[ (1-\mu) R(\mu) \Big] = 0 \eqend{.}
\end{equation}

Finally, using Eq.~\eqref{eq:resolvent_diff_gen_2} we obtain for antiperiodic boundary conditions that
\begin{splitequation}
\label{eq:resolvent_ns_diff}
&\lim_{\epsilon \to 0^+} \left[ R^\text{NS}_{ab}(\mu + \mathi \epsilon; x,y) - R^\text{NS}_{ab}(\mu - \mathi \epsilon; x,y) \right] \\
&\quad= \frac{1}{\mu (1-\mu)} \frac{\mathi}{L} \sinh\left[ \frac{1}{2} \Omega_a(x) - \frac{1}{2} \Omega_b(y) \right] \left( \frac{1}{\mu} - 1 \right)^{\frac{\mathi}{2 \pi} \left[ \Omega_a(x) - \Omega_b(y) \right]} \pf \frac{1}{\sin\left[ \frac{\pi}{L} (x-y) \right]} \delta_{ab} \\
&\quad= \frac{\mathi \sin\left( \frac{2 \pi \ell}{L} \right)}{2 L \sqrt{ \sin\left[ \frac{\pi}{L} (\ell+x) \right] \sin\left[ \frac{\pi}{L} (\ell-x) \right] \sin\left[ \frac{\pi}{L} (\ell+y) \right] \sin\left[ \frac{\pi}{L} (\ell-y) \right] }} \\
&\qquad\times \frac{1}{\mu (1-\mu)} \left( \frac{1}{\mu} - 1 \right)^{\frac{\mathi}{2 \pi} \left[ \Omega_a(x) - \Omega_b(y) \right]} \delta_{ab} \eqend{.} \raisetag{2em}
\end{splitequation}
Inserting these expressions into Eq.~\eqref{eq:flow_singleparticle_function_2}, it follows that
\begin{splitequation}
f_a(t,x) &= \int_{-\infty}^\infty \int_{-\ell}^\ell \frac{\sin\left( \frac{2 \pi \ell}{L} \right)}{2 L \sqrt{ \sin\left[ \frac{\pi}{L} (\ell+x) \right] \sin\left[ \frac{\pi}{L} (\ell-x) \right] \sin\left[ \frac{\pi}{L} (\ell+y) \right] \sin\left[ \frac{\pi}{L} (\ell-y) \right] }} \\
&\qquad\times \mathe^{- \mathi s \left[ \Omega_a(x) - \Omega_a(y) - 2 \pi t \right]} f_a(y) \total y \total s \eqend{,}
\end{splitequation}
where we changed variables to
\begin{equation}
\label{eq:flow_coordinate_change}
\mu = \frac{1}{1 + \mathe^{- 2 \pi s}} \eqend{,} \quad \frac{1}{\mu (1-\mu)} \total \mu = 2 \pi \total s \eqend{.}
\end{equation}
We can interchange the $y$ and $s$ integrals if we compute the $s$ integral in the distributional sense, which in this case is easy:
\begin{equation}
\int_{-\infty}^\infty \mathe^{- \mathi s \left[ \Omega_a(x) - \Omega_a(y) - 2 \pi t \right]} \total s = 2 \pi \delta\Big[ \Omega_a(x) - \Omega_a(y) - 2 \pi t \Big] \eqend{.}
\end{equation}
We thus obtain for antiperiodic boundary conditions that
\begin{equation}
f_a(t,x) = \sum_{b=1}^2 \int_{-\ell}^\ell K^\text{NS}_{ab}(t,x,y) f_b(y) \total y
\end{equation}
with the integral kernel
\begin{splitequation}
\label{eq:flow_kernel_ns_2}
K^\text{NS}_{ab}(t,x,y) &= \frac{\pi}{L} \frac{\sin\left( \frac{2 \pi \ell}{L} \right) \delta\Big[ \Omega_a(x) - \Omega_a(y) - 2 \pi t \Big] \delta_{ab}}{\sqrt{ \sin\left[ \frac{\pi}{L} (\ell+x) \right] \sin\left[ \frac{\pi}{L} (\ell-x) \right] \sin\left[ \frac{\pi}{L} (\ell+y) \right] \sin\left[ \frac{\pi}{L} (\ell-y) \right] }} \\
&= \frac{2 \pi}{L} \frac{\sinh\left[ \frac{1}{2} \Omega_1(x) - \frac{1}{2} \Omega_1(y) \right]}{\sin\left[ \frac{\pi}{L} (x-y) \right]} \delta\Big[ \Omega_a(x) - \Omega_a(y) - 2 \pi t \Big] \delta_{ab} \\
&= \frac{2 \pi}{L} \frac{\sinh(\pi t)}{\sin\left[ \frac{\pi}{L} (x-y) \right]} \delta\Big[ \Omega_a(x) - \Omega_a(y) - 2 \pi t \Big] \begin{pmatrix} 1 & 0 \\ 0 & -1 \end{pmatrix} \eqend{,}
\end{splitequation}
where we used the relation~\eqref{eq:resolvent_sinh_trafo}. We have thus proven the local flow for antiperiodic boundary conditions, Eq.~\eqref{eq:flow_kernel_ns} of Theorem~\ref{thm:flow}.

For periodic boundary conditions where the resolvent integral kernel is given in Eq.~\eqref{eq:resolvent_r_kernel_2}, the analogous computation using Eq.~\eqref{eq:resolvent_diff_gen_2} shows that
\begin{splitequation}
\label{eq:resolvent_r_diff}
&\lim_{\epsilon \to 0^+} \Big[ R^\text{R}_{ab}(\mu + \mathi \epsilon; x,y) - R^\text{R}_{ab}(\mu - \mathi \epsilon; x,y) \Big] \\
&\quad= \frac{\mathi \sin\left( \frac{2 \pi \ell}{L} \right) \cos\left[ \frac{\pi}{L} (x-y) \right]}{2 L \sqrt{ \sin\left[ \frac{\pi}{L} (\ell+x) \right] \sin\left[ \frac{\pi}{L} (\ell-x) \right] \sin\left[ \frac{\pi}{L} (\ell+y) \right] \sin\left[ \frac{\pi}{L} (\ell-y) \right] }} \\
&\qquad\times \frac{1}{\mu (1-\mu)} \left( \frac{1}{\mu} - 1 \right)^{\frac{\mathi}{2\pi} \left[ \Omega_a(x) - \Omega_b(y) \right]} \delta_{ab} \\
&\qquad+ \frac{1}{\mu (1-\mu)} \left( \frac{1}{\mu} - 1 \right)^{\frac{\mathi}{2\pi} \left[ \Omega_a(x) - \Omega_b(y) \right]} \left[ \mathe^{- \frac{1}{2} \left[ \Omega_a(x) - \Omega_b(y) \right]} g_{ab}^+(\mu) - \mathe^{\frac{1}{2} \left[ \Omega_a(x) - \Omega_b(y) \right]} g_{ab}^-(\mu) \right] \eqend{,}
\end{splitequation}
where we defined the boundary values of the matrix $g$~\eqref{eq:riemannhilbert_g_def} by
\begin{splitequation}
\label{eq:riemannhilbert_g_boundary}
g^\pm(\mu) &\equiv \lim_{\epsilon \to 0^+} g(\mu \pm \mathi \epsilon) \\
&= \frac{2 h - \tr h \, \1 + \sum_{\sigma=\pm 1} \sigma \frac{( 1 + 2 \sigma L h_1 ) ( 1 + 2 \sigma L h_2 )}{4 L} \left( \frac{1}{\mu} - 1 \right)^{\sigma \frac{2 \ell}{L}} \mathe^{\pm \sigma \frac{2 \pi \mathi \ell}{L}} \1}{1 - 4 L^2 h_1 h_2 + \sum_{\sigma=\pm 1} \frac{1}{2} ( 1 + 2 \sigma L h_1 ) ( 1 + 2 \sigma L h_2 ) \left( \frac{1}{\mu} - 1 \right)^{\sigma \frac{2 \ell}{L}} \mathe^{\pm \sigma \frac{2 \pi \mathi \ell}{L}}} \eqend{.}
\end{splitequation}
Clearly the first term of Eq.~\eqref{eq:resolvent_r_diff} is identical to the one for antiperiodic boundary conditions~\eqref{eq:resolvent_ns_diff}, such that also its contribution to the integral kernel of the modular flow will be identical to~\eqref{eq:flow_kernel_ns_2}. For the second term, we again change coordinates according to~\eqref{eq:flow_coordinate_change} in Eq.~\eqref{eq:flow_singleparticle_function_2}. We then notice that
\begin{equation}
g^\pm\left( \frac{1}{1 + \mathe^{- 2 \pi s}} \right) = g\left( \frac{1}{1 + \mathe^{- 2 \pi s \pm \mathi \pi}} \right) \eqend{,}
\end{equation}
such that the contribution of the second term to the integral kernel of the modular flow reads
\begin{splitequation}
\label{eq:flow_kernel_r_part1}
&- \mathi \int_{-\infty}^\infty \sum_{b=1}^2 \int_{-\ell}^\ell \left[ \mathe^{- \frac{1}{2} \left[ \Omega_a(x) - \Omega_b(y) \right]} g_{ab}\left( \frac{1}{1 + \mathe^{- 2 \pi s + \mathi \pi}} \right) - \mathe^{\frac{1}{2} \left[ \Omega_a(x) - \Omega_b(y) \right]} g_{ab}\left( \frac{1}{1 + \mathe^{- 2 \pi s - \mathi \pi}} \right) \right] \\
&\hspace{8em}\times \mathe^{- \mathi s \left[ \Omega_a(x) - \Omega_b(y) - 2 \pi t \right]} f_b(y) \total y \total s \\
&= - \mathi \int_{-\infty}^\infty \sum_{b=1}^2 \int_{-\ell}^\ell \bigg[ g_{ab}\left( \frac{1}{1 + \mathe^{- 2 \pi \left( s - \frac{\mathi}{2} \right)}} \right) \mathe^{- \mathi \left( s - \frac{\mathi}{2} \right) \left[ \Omega_a(x) - \Omega_b(y) - 2 \pi t \right]} \, \mathe^{- \pi t} \\
&\hspace{8em}- g_{ab}\left( \frac{1}{1 + \mathe^{- 2 \pi \left( s + \frac{\mathi}{2} \right)}} \right) \mathe^{- \mathi \left( s + \frac{\mathi}{2} \right) \left[ \Omega_a(x) - \Omega_b(y) - 2 \pi t \right]} \, \mathe^{\pi t} \bigg] f_b(y) \total y \total s \eqend{.} \raisetag{5.6em}
\end{splitequation}
Using the fact that $f_a \in L^2([-\ell,\ell])$ and the Cauchy--Schwarz inquality, we estimate
\begin{splitequation}
\abs{ \int_{-\ell}^\ell \mathe^{\mathi (s+\mathi t) \Omega_b(y)} f_b(y) \total y }^2 &\leq \int_{-\ell}^\ell \abs{\mathe^{\mathi (s+\mathi t) \Omega_b(y)} }^2 \total y \int_{-\ell}^\ell \abs{ f_b(y) }^2 \total y \\
&= \norm{ f_b }_2 \int_{-\ell}^\ell \left( \frac{\sin\left[ \frac{\pi}{L} (\ell+x) \right]}{\sin\left[ \frac{\pi}{L} (\ell-x) \right]} \right)^{2 t} \total x \eqend{.}
\end{splitequation}
This is finite as long as $2 \abs{t} < 1$, and in the limit $t \to \pm \frac{1}{2}$ we obtain
\begin{equation}
\int_{-\ell}^\ell \left( \frac{\sin\left[ \frac{\pi}{L} (\ell+x) \right]}{\sin\left[ \frac{\pi}{L} (\ell-x) \right]} \right)^{2 t} \total x = \frac{L \sin\left( \frac{2 \pi \ell}{L} \right)}{\pi ( 2 \abs{t} - 1 )} + \bigo{ ( 2 \abs{t} - 1 )^0 } \eqend{.}
\end{equation}
Since this diverges only polynomially, it follows that the integral $\int_{-\ell}^\ell \mathe^{\mathi s \Omega_b(y)} f_b(y) \total y$ is an analytic function for $\Im s \in \left( - \frac{1}{2}, \frac{1}{2} \right)$, and the boundary values as $\Im s \to \pm \frac{1}{2}$ exist as distributions.

We can thus shift $s \to s \pm \frac{\mathi}{2}$ in the integrals in Eq.~\eqref{eq:flow_kernel_r_part1} when computing them in the distributional sense, which reduces them to
\begin{splitequation}
\label{eq:flow_kernel_r_part2}
&2 \mathi \sinh(\pi t) \sum_{b=1}^2 \int_{-\ell}^\ell \int_{-\infty}^\infty g_{ab}\left( \frac{1}{1 + \mathe^{- 2 \pi s}} \right) \mathe^{- \mathi s \left[ \Omega_a(x) - \Omega_b(y) - 2 \pi t \right]} \total s \, f_b(y) \total y \\
&= 2 \mathi \sinh(\pi t) \sum_{b=1}^2 \int_{-\ell}^\ell \int_{-\infty}^\infty \frac{2 h_{ab} - \tr h \, \delta_{ab} + \sum_{\sigma=\pm 1} \sigma \frac{( 1 + 2 \sigma L h_1 ) ( 1 + 2 \sigma L h_2 )}{4 L} \mathe^{- \sigma \frac{4 \pi \ell}{L} s} \delta_{ab}}{1 - 4 L^2 h_1 h_2 + \sum_{\sigma=\pm 1} \frac{1}{2} ( 1 + 2 \sigma L h_1 ) ( 1 + 2 \sigma L h_2 ) \mathe^{- \sigma \frac{4 \pi \ell}{L} s}} \\
&\hspace{10em}\times \mathe^{- \mathi s \left[ \Omega_a(x) - \Omega_b(y) - 2 \pi t \right]} \total s f_b(y) \total y \eqend{.}
\end{splitequation}
To compute the integral over $s$, we need the result~\cite[Eq.~(B.17)]{cadamurofroebminz2023}
\begin{equation}
\label{eq:flow_integral}
\int_{-\infty}^\infty \frac{\mathe^{\mathi s z}}{c + \mathe^{2 \pi s}} \total s = - \frac{\mathi}{2 c} \lim_{\epsilon \to 0^+} \left[ c^\frac{\epsilon + \mathi z}{2 \pi} \frac{1}{\sinh\left( \frac{z - \mathi \epsilon}{2} \right)} \right] = c^{\frac{\mathi z}{2 \pi} - 1} \left[ - \frac{\mathi}{2} \pf \frac{1}{\sinh\left( \frac{z}{2} \right)} + \pi \delta(z) \right] \eqend{,}
\end{equation}
valid for $z \in \mathbb{R}$ and $c \in \mathbb{C}$ with $\Re c > 0$. Decomposing the integrand of Eq.~\eqref{eq:flow_kernel_r_part2} in partial fractions and rescaling the integration variable $s$ by a constant, all integrals can be brought into the form~\eqref{eq:flow_integral}, and Eq.~\eqref{eq:flow_kernel_r_part2} reduces to
\begin{splitequation}
\label{eq:flow_kernel_r_part3}
&\frac{\mathi \sinh(\pi t)}{2 \ell ( h_1 - h_2 )} \sum_{b=1}^2 \int_{-\ell}^\ell \sum_{i=1}^2 ( \delta_{i,1} - \delta_{i,2} ) \sum_{\sigma = \pm 1} \sigma \left[ h_{ab} + (h_i - \tr h) \, \delta_{ab} - \sigma \frac{1 - 4 L^2 h_1 h_2}{4 L} \, \delta_{ab} \right] \\
&\qquad\times \frac{( 1 + 2 \sigma L h_i )}{( 1 - 2 \sigma L h_i )} \int_{-\infty}^\infty \frac{\mathe^{- \mathi \sigma \frac{L}{2 \ell} s \left[ \Omega_a(x) - \Omega_b(y) - 2 \pi t \right]}}{\mathe^{2 \pi s} + \frac{( 1 + 2 \sigma L h_i )}{( 1 - 2 \sigma L h_i )}} \total s f_b(y) \total y \\
&= \frac{\sinh(\pi t)}{4 \ell ( h_1 - h_2 )} \sum_{b=1}^2 \int_{-\ell}^\ell \sum_{i=1}^2 ( \delta_{i,1} - \delta_{i,2} ) \sum_{\sigma = \pm 1} \left[ h_{ab} + (h_i - \tr h) \, \delta_{ab} - \sigma \frac{1 - 4 L^2 h_1 h_2}{4 L} \, \delta_{ab} \right] \\
&\qquad\times \Bigg[ - \left[ \frac{( 1 + 2 L h_i )}{( 1 - 2 L h_i )} \right]^{- \mathi \frac{L}{4 \pi \ell} \left[ \Omega_a(x) - \Omega_b(y) - 2 \pi t \right]} \pf \frac{1}{\sinh\left[ \frac{L}{4 \ell} \left[ \Omega_a(x) - \Omega_b(y) - 2 \pi t \right] \right]} \\
&\qquad\quad+ 2 \pi \mathi \sigma \, \delta\left[ \frac{L}{2 \ell} \left[ \Omega_a(x) - \Omega_b(y) - 2 \pi t \right] \right] \Bigg] f_b(y) \total y \eqend{.} \raisetag{2.4em}
\end{splitequation}
Combining this result with the local terms (which as we found are identical to the antiperiodic case), we thus obtain for periodic boundary conditions that
\begin{equation}
f_a(t,x) = \sum_{b=1}^2 \int_{-\ell}^\ell K^\text{R}_{ab}(t,x,y) f_b(y) \total y
\end{equation}
with the integral kernel
\begin{splitequation}
\label{eq:flow_kernel_r_2}
K^\text{R}_{ab}(t,x,y) &= K^\text{NS}_{ab}(t,x,y) + \frac{1}{4 \ell} \sinh(\pi t) \, \pf \frac{1}{\sinh\left[ \frac{L}{4 \ell} \big[ 2 \pi t - \Omega_a(x) + \Omega_b(y) \big] \right]} \\
&\qquad\times \Bigg[ \left( \frac{1 + 2 L h_1}{1 - 2 L h_1} \right)^{\mathi \frac{\big[ 2 \pi t - \Omega_a(x) + \Omega_b(y) \big] L}{4 \pi \ell}} \begin{pmatrix} 1 + \cos \psi & \sin \psi \, \mathe^{\mathi \phi} \\ \sin \psi \, \mathe^{- \mathi \phi} & 1 - \cos \psi \end{pmatrix}_{ab} \\
&\qquad\quad+ \left( \frac{1 + 2 L h_2}{1 - 2 L h_2} \right)^\frac{\mathi \big[ 2 \pi t - \Omega_a(x) + \Omega_b(y) \big] L}{4 \pi \ell} \begin{pmatrix} 1 - \cos \psi & - \sin \psi \, \mathe^{\mathi \phi} \\ - \sin \psi \, \mathe^{- \mathi \phi} & 1 + \cos \psi \end{pmatrix}_{ab} \Bigg] \eqend{.}
\end{splitequation}
We have thus proven the non-local flow for periodic boundary conditions, Eq.~\eqref{eq:flow_kernel_r} of Theorem~\ref{thm:flow}. Corollary~\ref{cor:flow} is an immediate consequence of Theorem~\ref{thm:flow} and Lemma~\ref{lemma:limit}, and we omit the straightforward computations.

It remains to determine the generator of the modular flow, the modular Hamiltonian, and we start again with antiperiodic boundary conditions where the integral kernel of the modular flow is given in Eq.~\eqref{eq:flow_kernel_ns_2}. Expanding the first line of that equation in $t$ after using the well-known composition formula for the Dirac $\delta$, we obtain
\begin{splitequation}
\label{eq:hamiltonian_kernel_ns_2}
K^\text{NS}_{ab}(t,x,y) &= \sqrt{ \frac{\sin\left[ \frac{\pi}{L} (\ell+x) \right] \sin\left[ \frac{\pi}{L} (\ell-x) \right]}{\sin\left[ \frac{\pi}{L} (\ell+y) \right] \sin\left[ \frac{\pi}{L} (\ell-y) \right] }} \begin{pmatrix} \delta(x - x_0(y,t)) & 0 \\ 0 & \delta(x - x_0(y,-t)) \end{pmatrix} \\
&= \delta_{ab} \delta(x-y) - t \delta'(x-y) \frac{2 L}{\sin\left( \frac{2 \pi \ell}{L} \right)} \begin{pmatrix} 1 & 0 \\ 0 & - 1 \end{pmatrix}_{ab} \\
&\qquad\times \sqrt{ \frac{\sin\left[ \frac{\pi}{L} (\ell+x) \right] \sin\left[ \frac{\pi}{L} (\ell-x) \right]}{\sin\left[ \frac{\pi}{L} (\ell+y) \right] \sin\left[ \frac{\pi}{L} (\ell-y) \right] }} \left[ \sin^2\left( \frac{\pi \ell}{L} \right) - \sin^2\left( \frac{\pi y}{L} \right) \right] + \bigo{t^2} \eqend{,}
\end{splitequation}
where we used that the unique solution of $\Omega_1(x) - \Omega_1(y) - 2 \pi t = 0$ for $x \in [-\ell,\ell]$ reads
\begin{splitequation}
x_0(y,t) &\equiv \frac{L}{\pi} \arctan\left[ \tan\left( \frac{\pi \ell}{L} \right) \tanh\left( \frac{\Omega_1(y)}{2} + \pi t \right) \right] \\
&= \frac{L}{\pi} \arctan\left[ \tan\left( \frac{\pi \ell}{L} \right) \frac{\sin\left[ \frac{\pi}{L} (\ell+y) \right] \mathe^{2 \pi t} - \sin\left[ \frac{\pi}{L} (\ell-y) \right]}{\sin\left[ \frac{\pi}{L} (\ell+y) \right] \mathe^{2 \pi t} + \sin\left[ \frac{\pi}{L} (\ell-y) \right]} \right] \\
&= y + \frac{2 L}{\sin\left( \frac{2 \pi \ell}{L} \right)} \left[ \sin^2\left( \frac{\pi \ell}{L} \right) - \sin^2\left( \frac{\pi y}{L} \right) \right] t + \bigo{t^2} \eqend{.}
\end{splitequation}
Employing the relation $K = \mathe^{\mathi t H}$ between the integral kernels of the modular flow $K$ and the modular Hamiltonian $H$, which reads concretely
\begin{equation}
K^\text{NS}_{ab}(t,x,y) = \delta_{ab} \delta(x-y) + \mathi t H^\text{NS}_{ab}(x,y) + \bigo{t^2} \eqend{,}
\end{equation}
we can then read off the latter kernel:
\begin{splitequation}
H^\text{NS}_{ab}(x,y) &= \mathi \delta'(x-y) \frac{2 L}{\sin\left( \frac{2 \pi \ell}{L} \right)} \begin{pmatrix} 1 & 0 \\ 0 & - 1 \end{pmatrix}_{ab} \\
&\qquad\times \sqrt{ \frac{\sin\left[ \frac{\pi}{L} (\ell+x) \right] \sin\left[ \frac{\pi}{L} (\ell-x) \right]}{\sin\left[ \frac{\pi}{L} (\ell+y) \right] \sin\left[ \frac{\pi}{L} (\ell-y) \right] }} \left[ \sin^2\left( \frac{\pi \ell}{L} \right) - \sin^2\left( \frac{\pi y}{L} \right) \right] \\
&= \mathi \delta'(x-y) \frac{2 L}{\sin\left( \frac{2 \pi \ell}{L} \right)} \begin{pmatrix} 1 & 0 \\ 0 & - 1 \end{pmatrix}_{ab} \left[ \sin^2\left( \frac{\pi \ell}{L} \right) - \sin\left( \frac{\pi x}{L} \right) \sin\left( \frac{\pi y}{L} \right) \right] \eqend{.}
\end{splitequation}
We have thus proven Eq.~\eqref{eq:hamiltonian_kernel_ns} of Theorem~\ref{thm:hamiltonian}.

For antiperiodic boundary conditions, the expansion of the modular flow kernel~\eqref{eq:flow_kernel_r_2} is now straightforward, and we obtain Eq.~\eqref{eq:hamiltonian_kernel_r} of Theorem~\ref{thm:hamiltonian}. Corollary~\ref{cor:hamiltonian} is an immediate consequence of Theorem~\ref{thm:hamiltonian} and Lemma~\ref{lemma:limit}, and we omit the straightforward computations.

\section{Outlook}
\label{sec:outlook}

We have considered the most general family of quasi-free zero-energy ground states that exist for fermions on a cylinder with periodic boundary conditions, and determined the modular flow for this state and the algebra generated by fermions restricted to an interval. Our results greatly generalize existing results, and in general, the flow (and its generator, the modular Hamiltonian) is a non-local operator that mixes both chiralities. Our results for the modular flow and the modular Hamiltonian can now be employed, for example, to determine quantities such as the relative entropy between an exited state and the ground state, analogously to what has been done in other situations~\cite{longo2019,casinigrillopontello2019,hollands2019,galandamuchverch2023,froebmuchpapadopoulos2023,dangeloetal2024}.

In contrast to the case of antiperiodic boundary conditions (see for example~\cite{rehrentedesco2013}), it is not clear if our results could also be obtained by a conformal mapping from Minkowski spacetime. Since in flat space the modular Hamiltonian can be written as a linear combination of conformal Noether charges, a straightforward mapping would imply that also for the cylinder this holds true, and in fact one would expect exactly the same linear combination. In fact, for antiperiodic boundary conditions a short computation using the canonical anticommutation relations~\eqref{eq:car} readily establishes that $[ \ln \Delta, \psi(f) ] = [ Q^\text{NS}, \psi(f) ]$ with the conformal Noether charge
\begin{equation}
Q^\text{NS} = \int_0^L T_{\mu\nu}(0,x) \xi^\mu(0,x) n^\nu \total x
\end{equation}
evaluated on the Cauchy surface $t = 0$ with normal vector $n^\mu = \delta^\mu_0$. The fermionic stress tensor $T^{\mu\nu}$ is the usual (normal-ordered) one
\begin{equation}
T_{\mu\nu} = \frac{1}{2} \normord{ \bar\psi \gamma_{(\mu} \nabla_{\nu)} \psi } - \frac{1}{2} \normord{ \nabla_{(\mu} \bar\psi \gamma_{\nu)} \psi } \eqend{,}
\end{equation}
and the conformal Killing vector $\xi^\mu$ has the components
\begin{equations}
\xi^0(t,x) &= - \frac{L}{2} \frac{2 \cos\left( \frac{2 \pi \ell}{L} \right) - \cos\left[ \frac{2 \pi (x+t)}{L} \right] - \cos\left[ \frac{2 \pi (x-t)}{L} \right]}{\sin\left( \frac{2 \pi \ell}{L} \right)} \eqend{,} \\
\xi^1(t,x) &= - L \frac{\sin\left( \frac{2 \pi x}{L} \right) \sin\left( \frac{2 \pi t}{L} \right)}{\sin\left( \frac{2 \pi \ell}{L} \right)} \eqend{.}
\end{equations}
In the limit $L \to \infty$, this reduces to the known flat-space result.

However, for periodic boundary conditions the corresponding link is much more difficult, even for pure states. Let us consider the states with $h_1 = h_2 = \pm \frac{1}{2 L}$, where the integral kernel of the modular Hamiltonian is given by Eq.~\eqref{eq:hamiltonian_limit_eq}, and where the two chiralities of the fermion are still uncoupled. The additional term proportional to $\delta(x-y)$ can be obtained from a commutator of the fermion field with the current $j^\mu = \mathi \normord{ \bar\psi \gamma^\mu \psi }$, integrated over the Cauchy surface $t = 0$ with the smearing function $\xi(x) = 2 \pi \frac{\sin^2\left( \frac{\pi}{L} \ell \right) - \sin^2\left( \frac{\pi}{L} x \right)}{\sin\left( \frac{2 \pi \ell}{L} \right)}$. We thus obtain $[ \ln \Delta, \psi(f) ] = [ Q^\text{R}_\pm, \psi(f) ]$ with
\begin{equation}
Q^\text{R}_\pm = \int_0^L \Big[ T_{\mu\nu}(0,x) \xi^\mu(0,x) \pm j_\nu(0,x) \xi(x) \Big] n^\nu \total x \eqend{,}
\end{equation}
but it is impossible to extend the integrand as a function of $t$ and $x$ in such a way that $Q^\text{R}$ is conserved, $\partial_t Q^\text{R} = 0$. Nevertheless, considering fermions of only one chirality, the quantities
\begin{equation}
Q^\text{R}_{s,\sigma}(t) = \int_0^L \Big[ T_{\mu\nu}(t,x) \xi^\mu(t,x) + s j_\nu(t,x) \xi(x+\sigma t) \Big] n^\nu \total x
\end{equation}
fulfill
\begin{equation}
\partial_t [ Q^\text{R}_{s,+}(t), \psi_1(y) ] = 0 = \partial_t [ Q^\text{R}_{s,-}(t), \psi_2(y) ] \eqend{.}
\end{equation}
That is, while the ``charges'' themselves are not conserved, their commutator with one component of the free fermion field is time-independent. This result hinges on the fact that each component is a function of a single null coordinate $x \pm t$, and consequently the commutator with the other component will depend on time.

For a general ground state which couples the two chiralities, the problem will only be exacerbated. Even for a local flow, namely the states with $h_1 = - h_2 = \pm \frac{1}{2 L}$ where the integral kernel of the modular Hamiltonian is given by Eq.~\eqref{eq:hamiltonian_limit_neq}, the chiralities can be coupled, and so our results seem to be in contradiction with the arguments of~\cite{sorce2024}. It will be very interesting to study these questions further and to determine the exact conditions under which a local and geometric flow arises, as well as to generalize our results to multiple intervals or thermal states.

\begin{acknowledgments}
This work has been funded by the Deutsche Forschungsgemeinschaft (DFG, German Research Foundation) --- project no. 396692871 within the Emmy Noether grant CA1850/1-1. M. B. F. thanks G. P.-N. and the Grupo de Física Teórica de Altas Energías of the Universidad de Buenos Aires for their hospitality.
\end{acknowledgments}

\appendix

\section{Limits of distributions}
\label{app:lemma_limit}

We consider the distribution 
\begin{equation}
\label{eq:app_lemma_limit_distr}
a^{\mathi t} \, \pf \frac{1}{\sinh\left( \pi t \right)} \eqend{,}
\end{equation}
which acts on a continuous function $f$ that satisfies $\int_{-\infty}^\infty \abs{ \frac{f(t)}{\sinh(\pi t)} } \total t < \infty$ according to
\begin{equation}
\label{eq:app_lemma_limit_action}
\int a^{\mathi t} \, \pf \frac{1}{\sinh\left( \pi t \right)} f(t) \total t \equiv \lim_{\epsilon \to 0^+} \left[ \int_{-\infty}^{-\epsilon} \frac{a^{\mathi t}}{\sinh\left( \pi t \right)} f(t) \total t + \int_\epsilon^\infty \frac{a^{\mathi t}}{\sinh\left( \pi t \right)} f(t) \total t \right] \eqend{.}
\end{equation}
In the integrand, we add and subtract $f(0)$ such that
\begin{splitequation}
\label{eq:app_lemma_limit_action2}
\int a^{\mathi t} \, \pf \frac{1}{\sinh\left( \pi t \right)} f(t) \total t &= \int_{-\infty}^\infty \frac{a^{\mathi t}}{\sinh\left( \pi t \right)} [ f(t) - f(0) ] \total t \\
&\quad+ f(0) \lim_{\epsilon \to 0^+} \int_\epsilon^\infty \frac{a^{\mathi t} - a^{- \mathi t}}{\sinh\left( \pi t \right)} \total t \eqend{.}
\end{splitequation}
We thus have to compute the integral
\begin{equation}
\int_\epsilon^\infty \frac{a^{\mathi t}}{\sinh\left( \pi t \right)} \total t = \frac{a^{\mathi \epsilon} \mathe^{\pi \epsilon}}{\pi} \int_0^1 \frac{u^{- \frac{1}{2} - \frac{\mathi}{2 \pi} \ln a}}{\mathe^{2 \pi \epsilon} - u} \total u \eqend{,}
\end{equation}
where we made a change of variables to $u = \mathe^{- 2 \pi (t-\epsilon)}$. As $\epsilon \to 0$, the integrand develops a singularity at $u = 1$, which we isolate by writing
\begin{splitequation}
\int_0^1 \frac{u^{- \frac{1}{2} - \frac{\mathi}{2 \pi} \ln a}}{\mathe^{2 \pi \epsilon} - u} \total u &= \int_0^1 \frac{1}{\mathe^{2 \pi \epsilon} - u} \total u + \int_0^1 \frac{u^{- \frac{1}{2} - \frac{\mathi}{2 \pi} \ln a} - 1}{\mathe^{2 \pi \epsilon} - u} \total u \\
&= \ln\left( \frac{\mathe^{2 \pi \epsilon}}{\mathe^{2 \pi \epsilon} - 1} \right) + \int_0^1 \frac{u^{- \frac{1}{2} - \frac{\mathi}{2 \pi} \ln a} - 1}{1 - u} \total u + \bigo{\epsilon} \\
&= - \ln(2 \pi \epsilon) - \gamma - \psi\left( \frac{1}{2} - \frac{\mathi}{2 \pi} \ln a \right) + \bigo{\epsilon} \eqend{,}
\end{splitequation}
and where in the last step we used the integral representation~\cite[Eq.~(5.9.16)]{dlmf} of the digamma function. It follows that the action~\eqref{eq:app_lemma_limit_action2} can be written as
\begin{equation}
\label{eq:app_lemma_limit_action3}
\int a^{\mathi t} \, \pf \frac{1}{\sinh\left( \pi t \right)} f(t) \total t = \int_{-\infty}^\infty \frac{a^{\mathi t}}{\sinh\left( \pi t \right)} [ f(t) - f(0) ] \total t + \mathi f(0) \frac{a-1}{a+1} \eqend{,}
\end{equation}
where we also used the reflection formula~\cite[Eq.~(5.5.4)]{dlmf} for the digamma function.

By assumption, we have $g \in L^1(\mathbb{R})$ for the function
\begin{equation}
g(t) \equiv \frac{f(t) - f(0)}{\sinh\left( \pi t \right)} \eqend{,}
\end{equation}
such that the Riemann--Lebesgue lemma is applicable to $g$. We thus obtain from Eq.~\eqref{eq:app_lemma_limit_action3} the two limits
\begin{equations}
\lim_{a \to 0} \int \left[ a^{\mathi t} \, \pf \frac{1}{\sinh\left( \pi t \right)} - \mathi \frac{a-1}{a+1} \delta(t) \right] f(t) \total t &= 0 \eqend{,} \\
\lim_{a \to \infty} \int \left[ a^{\mathi t} \, \pf \frac{1}{\sinh\left( \pi t \right)} - \mathi \frac{a-1}{a+1} \delta(t) \right] f(t) \total t &= 0 \eqend{,}
\end{equations}
which proves Lemma~\ref{lemma:limit}.

\bibliography{references}

\end{document}